%% file: 0_main.tex
\documentclass[sigconf]{acmart}

\AtBeginDocument{%
  }

\usepackage{multirow}
\usepackage{enumitem}
\usepackage{colortbl}
\usepackage{tablefootnote}
\usepackage{makecell}
\usepackage{balance}

\newcommand{\ie}{\emph{i.e., }}
\newcommand{\eg}{\emph{e.g., }}
\newcommand{\etal}{\emph{et al. }}

\newcommand{\wrt}{\emph{w.r.t. }}

\copyrightyear{2024}
\acmYear{2024}
\setcopyright{acmlicensed}\acmConference[MM '24]{Proceedings of the 32nd ACM International Conference on Multimedia}{October 28-November 1, 2024}{Melbourne, VIC, Australia}
\acmBooktitle{Proceedings of the 32nd ACM International Conference on Multimedia (MM '24), October 28-November 1, 2024, Melbourne, VIC, Australia}
\acmDOI{10.1145/3664647.3681593}
\acmISBN{979-8-4007-0686-8/24/10}

\begin{document}

\title{MM-Forecast: A Multimodal Approach to Temporal Event Forecasting with Large Language Models}

\author{Haoxuan Li}
\affiliation{%
  \institution{University of Electronic Science and Technology of China}
  \city{Chengdu}
  \country{China}
}
\email{lhx980610@gmail.com}

\author{Zhengmao Yang}
\affiliation{%
  \institution{Zhejiang University}
  \city{Hangzhou}
  \country{China}
}
\email{zmyang4671@zju.edu.cn}

\author{Yunshan Ma}
\affiliation{%
  \institution{National University of Singapore}
  \country{Singapore}
}
\email{yunshan.ma@u.nus.edu}

\author{Yi Bin}
\authornote{Yi Bin is the corresponding author (Email: yi.bin@hotmail.com)}
\affiliation{%
  \institution{Tongji University}
  \city{Shanghai}
  \country{China}
}
\affiliation{
  \institution{National University of Singapore}
  \country{Singapore}
}
\email{yi.bin@hotmail.com}

\author{Yang Yang}
\affiliation{%
  \institution{University of Electronic Science and Technology of China}
  \city{Chengdu}
  \country{China}
}
\email{yang.yang@uestc.edu.cn}

\author{Tat-Seng Chua}
\affiliation{%
  \institution{National University of Singapore}
  \country{Singapore}
}
\email{dcscts@nus.edu.sg}

\renewcommand{\shortauthors}{Haoxuan Li et al.}

\begin{abstract}
  We study an emerging and intriguing problem of multimodal temporal event forecasting with large language models. Compared to using text or graph modalities, the investigation of utilizing images for temporal event forecasting has not been fully explored, especially in the era of large language models (LLMs). To bridge this gap, we are particularly interested in two key questions of: 1) why images will help in temporal event forecasting, and 2) how to integrate images into the LLM-based forecasting framework. To answer these research questions, we propose to identify two essential functions that images play in the scenario of temporal event forecasting, \ie \textbf{highlighting and complementary}. Then, we develop a novel framework, named \textbf{MM-Forecast}. It employs an Image Function Identification module to recognize these functions as verbal descriptions using multimodal large language models (MLLMs), and subsequently incorporates these function descriptions into LLM-based forecasting models. To evaluate our approach, we construct a new multimodal dataset, MidEast-TE-mm, by extending an existing event dataset MidEast-TE-mini with images. Empirical studies demonstrate that our MM-Forecast can correctly identify the image functions, and further more, incorporating these verbal function descriptions significantly improves the forecasting performance. 
  The dataset, code, and prompts are available at \textcolor{magenta}{{\url{https://github.com/LuminosityX/MM-Forecast}}}.
\end{abstract}




\maketitle

\input{1_introduction}
\input{2_related_works}

\input{3_method}
\input{4_experiments}

\input{5_conclusion}

\input{6_ack}


\bibliographystyle{ACM-Reference-Format}
\balance
\bibliography{0_main}

\clearpage
\appendix

\input{7_appendix}

\end{document}

%% file: 1_introduction.tex
\section{Introduction}

Temporal event forecasting aims to predict future events according the observed events in history. The forecasting of critical events, such as pandemic outbreak, civil unrest, and international conflicts, can help shape policies in advance and minimize potential impacts~\cite{zhao2021event}. Due to its great potential application value, temporal event forecasting~\cite{ForecastQA, LoGo, scriptLearning, Foundation_TS} has garnered increasing attention from both the academic and industrial community. Despite promising progress, current methods have ignored the rich multimodal information, \eg images, leaving this an unexplored research gap.

With the enormous success of LLMs, an increasing number of studies~\cite{GPT-NeoX-ICL, GENTKG, Chain-of-History, PPT, zhang2024analyzing, ye2024mirai, chang2024comprehensive} have been exploring LLMs to tackle the temporal event forecasting problem.
These pioneering works explore the application of LLMs in the task of temporal event forecasting, leveraging techniques such as in-context learning (ICL)~\cite{GPT-NeoX-ICL, zhang2024analyzing}, instruction tuning~\cite{Chain-of-History, PPT}, and retrieval-augmented generation (RAG)~\cite{ToG, chang2024comprehensive}. 
Compared to LLM-based methods, traditional methods have several shortcomings in terms of effectiveness, flexibility, and scalability. Specifically, traditional non-LLM methods~\cite{RENET, REGCN, LoGo, EvoKG, SeCoGD}, whether based on structured or unstructured data, typically require large-scale well-annotated datasets.
Moreover, model selection is often a challenge for these traditional methods due to high computational costs. Additionally, traditional methods generally require separate training for different datasets, as a result, they often struggle to make fast adaptation \wrt frequent changing in dataset and temporal shifts. 
Therefore, applying LLMs to the task of temporal event forecasting is a worthwhile direction to explore~\cite{GPT-NeoX-ICL}.
However, all of the existing LLM-based methods only consider a single modality, such as text~\cite{GPT-NeoX-ICL} or graph~\cite{Chain-of-History}, while ignoring the prevalent visual modality, \ie images. Some previous works~\cite{li2024focusing, li2023your} have justified that images are helpful in multimodal event detection~\cite{li2024focusing} and extraction~\cite{DRMM,CLIP-Event}, while none of them investigate images' utility in temporal event forecasting.
\begin{figure}[t] \label{fig:motivation}
  \centering
  \includegraphics[width=0.9\linewidth]{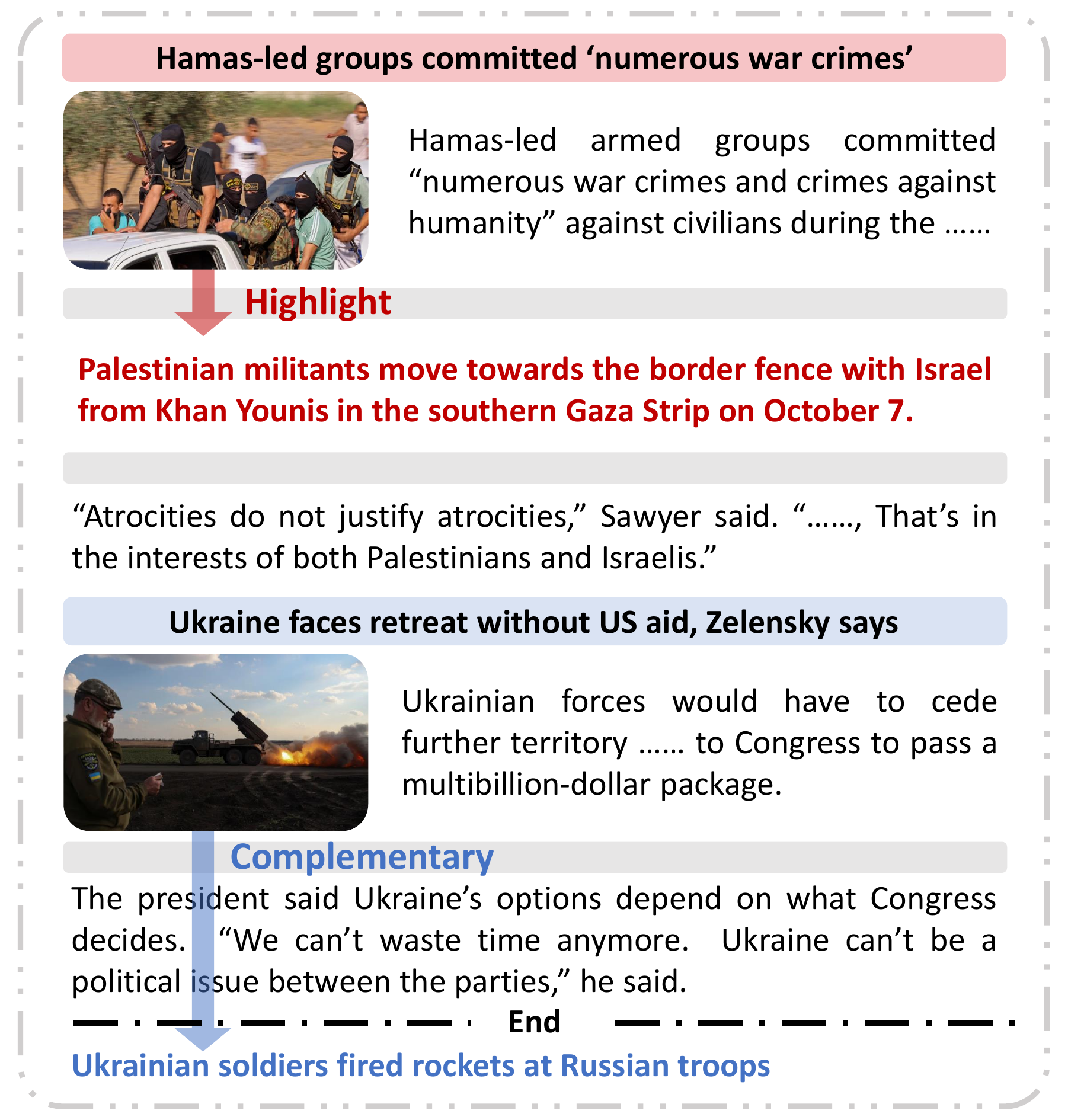}
  \vspace{-0.1in}
  \caption{Illustration of our motivation about why images will help in temporal event forecasting. We identify two essential functions of images, \ie highlighting and complementary. By offering auxiliary highlighting or complementary information, images enhance the understanding of temporal events, thus boosting the forecasting performance.}
\label{fig:motivation}
\vspace{-0.1in}
\end{figure}

To bridge this gap, we aim to integrate images into temporal event forecasting and construct multimodal temporal event forecasting models. However, it is a non-trivial objective due to the following challenges. 
First, it is necessary to clarify the function between visual information and other modal information, \ie the interplay between visual and non-visual modalities. Next, we need to figure out how the function between two modalities can contribute to the task of temporal event forecasting. Second, previous works~\cite{DRMM} that explores the image function typically require large amounts of labeled training data. Additionally, images serve different functions for different specific tasks, so these methods often struggle to generalize effectively to other task definitions.
Therefore, there is a pressing need to design an effective method to identify the function between modalities and seamlessly integrating them into LLM-based forecasting models.


To address the aforementioned challenges, we propose a novel framework for multimodal temporal event forecasting, named as \textbf{MM-Forecast}. Specifically, we identify two essential functions of images, \ie highlighting and complementary. As illustrated in Figure~\ref{fig:motivation}, when the function of associated image is highlighting, the image plays the role of emphasizing the key events. In contrast, when the function of associated image is complementary, the image provides supplementary information that complements the textual content. In order to recognize these two types of functions, we propose an Image Function Identification module that is based on Multimodal LLMs (MLLMs) due to their superior multimodal understanding and reasoning capabilities in zero-shot settings~\cite{li2024multimodal}. This proposed module is designed to recognize the function of images in historical events, and then transform this information into verbal descriptions that can be seamlessly integrated into the LLM-based event forecasting model. 
Equipping this Image Function Identification module into the overall framework, we integrate it into two distinct LLM-based forecasting models, \ie one based on the in-context learning (ICL) method~\cite{GPT-NeoX-ICL}, and the other based on the retrieval-augmented generation (RAG) technique~\cite{RAG}. 
In order to evaluate our approach, we construct an exploratory dataset by incorporating images into an existing dataset MidEast-TE-mini~\cite{chang2024comprehensive}. We name this new dataset MidEast-TE-multimodal (short as \textbf{MidEast-TE-mm}). 
In the final evaluation, with the enhancement of visual information, the temporal event forecasting task achieves superior forecasting accuracy compared to the unimodal approach. The experimental results illustrate that our method accurately recognizes the function of images in various aspects. Furthermore, the findings demonstrate that multimodal temporal forecasting represents a potential and promising research direction worthy of further exploration.
The main contributions are as follows:
\begin{itemize}[leftmargin=*]
    \item To the best of our knowledge, this is the first comprehensive study of exploring visual information for temporal event forecasting in the era of LLMs.
    \item We identify two main functions that images play in temporal event forecasting, and design a framework to recognise and integrate visual information into LLM-based forecasting models.
    \item Extensive experiments justify that our framework is able to identify the functions of images and visual information can enhance the performance of temporal event forecasting. Furthermore, these findings have led to several noteworthy and promising directions for future research.
\end{itemize}

%% file: 2_related_works.tex
\section{Related Works}
We survey the related works of temporal event forecasting and LLMs for event analysis. 

\subsection{Temporal Event Forecasting}
Temporal event forecasting centers on predicting future event occurrences based on historical events, and the typical approaches can be categorized by event format: time series, structured, and unstructured events. 
Regarding the time series paradigm, existing works~\cite{Foundation_TS, benjamin2023hybrid, RCT_B} typically represent events as an ordered sequence of data points that describe the progression of actions or occurrences. However, this paradigm inherently fails to represent multiple relationships among entities. Alternatively, another branch of works~\cite{ConvTransE, ConvE, RotatE, DistMult, SeCoGD, RENET, REGCN, EvoKG} focus on the prediction of structured events, \ie using graph to represent events, which is known as temporal knowledge graph (TKG).
Recent works\cite{SeCoGD, LoGo} introduce context into temporal event forecasting models, enhancing the prediction performance by elaborating the event’s occurrence situation. 
In addition, several studies have explored the use of unstructured textual representations of temporal events, where each atomic event is generated from multi-document summaries~\cite{gholipour-ghalandari-etal-2020-large} or event chains~\cite{eventChain}. 
Nonetheless, all of them design forecasting models relying on single modality data. Some works~\cite{DRMM,CLIP-Event} explore the image utility in event extraction task, while none of them investigate images’ utility in temporal event forecasting.

\subsection{LLMs for Event Analysis}
The tremendous success of LLMs in recent years, exemplified by ChatGPT and its numerous successors~\cite{LLaMA, opt, palm, vicuna2023}, has inspired researchers to explore the application of these powerful models to various event-related tasks~\cite{deng2024advances, GENTKG, GPT-NeoX-ICL, zhang2024analyzing, ye2024mirai, chang2024comprehensive}. 
One area of research focuses on temporal understanding, where LLMs are tested for the task of temporal event ordering or storyline understanding~\cite{torque, situatedqa, going}. 
More works focus on leveraging LLMs to tackle the typical task of temporal reasoning~\cite{tempbench, tram}, while the task of forecasting receives much less attention.
Deng \etal~\cite{deng2024advances} surveyed the recent advances in event modeling, ranging from graph neural networks to LLMs. Specifically, GENTKG~\cite{GENTKG} improves the selection of historical event inputs by a temporal logical rule-based retrieval strategy.
Beyond specific methods, more works are focusing on benchmarking LLMs' capability in temporal event forecasting. Zhang \etal~\cite{zhang2024analyzing} propose a method to evaluate the proficiency of LLMs in handling temporal dynamics and understanding extensive text through three distinct tasks. And Ye \etal~\cite{ye2024mirai} introduce a novel benchmarking environment designed to rigorously assess and advance the capabilities of LLM agents for international event forecasting over time. Furthermore, Chang \etal~\cite{chang2024comprehensive} propose a unified dataset of structured and unstructured data, and systematically evaluates LLM-based methods on the task of text-involved temporal event forecasting.
However, these existing LLM-based methods still solely rely on single-modality data, potentially missing valuable information from other modalities, such as images.
With the success of LLMs, MLLMs, such as LLaVA~\cite{llava}, and Gemini~\cite{gemini}, have emerged as promising means for unifying visual and textual modalities. 
These MLLMs have demonstrated impressive performance gain across various visual-language tasks~\cite{Flamingo, llava, bin2023unifying, ding2024fashionregen}, suggesting their potential in the task of temporal event forecasting by leveraging visual information.

%% file: 3_method.tex
\section{Our Approach: MM-Forecast}
\begin{figure*}
\centering
\includegraphics[width=0.9\textwidth]{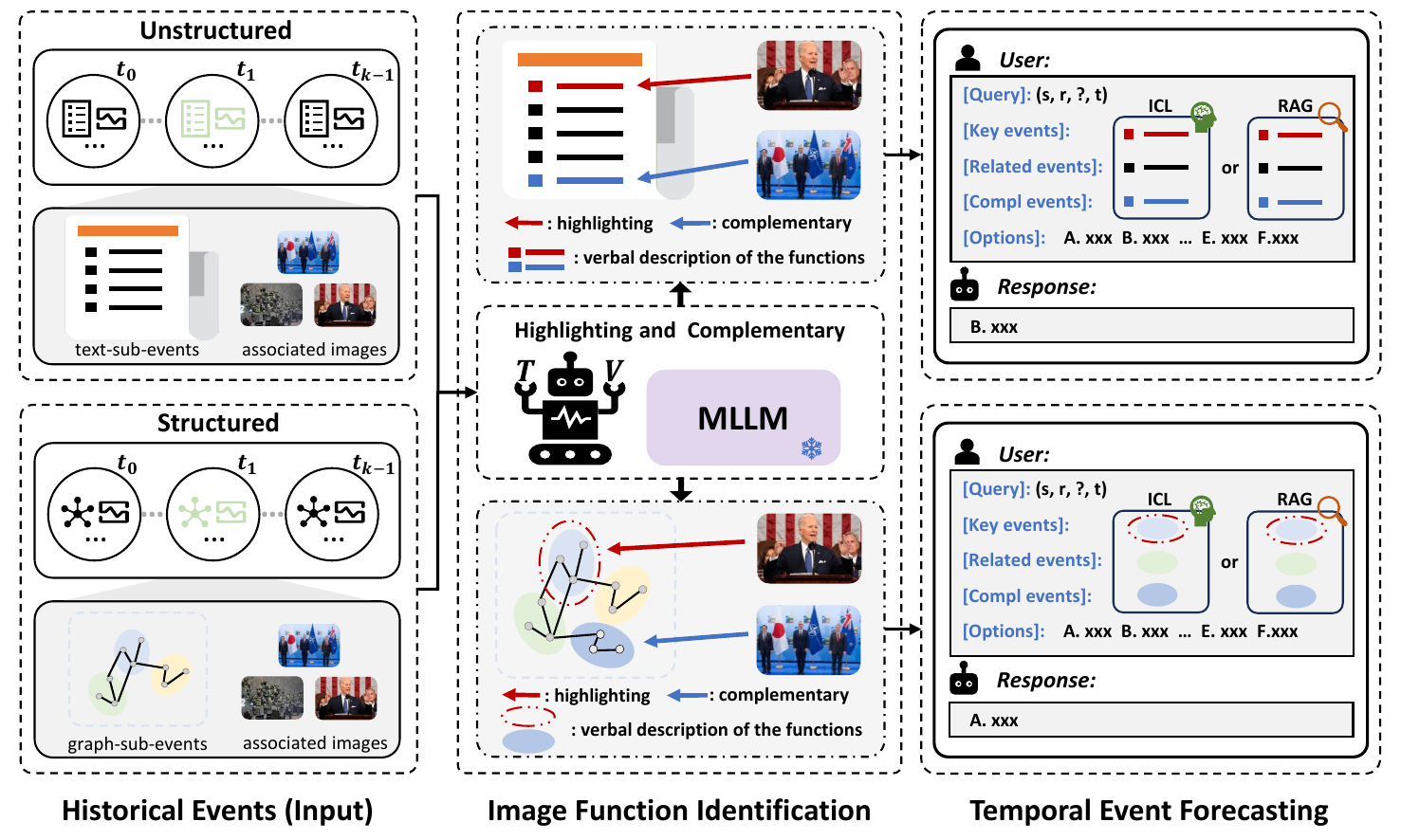}
\caption{The schematic overview of MM-Forecast. 
By consuming historical events in either format of unstructured or structured input (left), our image function identification module (middle) recognizes the image functions as verbal descriptions, which are then feed into LLM-based forecasting model (right).
Our framework is versatile to handle both structured and unstructured events, meanwhile, it is compatible to popular LLM components for event forecasting, \ie ICL and RAG.
}
\label{fig:overall_framework}
\end{figure*}

The overall framework of our proposed approach is depicted in Figure~\ref{fig:overall_framework}. We first formally define the multimodal temporal event forecasting task in Section~\ref{sec:3_1}. Second, we specifically introduce the key module of Image Function Identification in Section~\ref{sec:3_2}. Finally, we elaborate on how to integrate the recognized image functions into LLM-based forecasting models in Section~\ref{sec:3_3}.

\subsection{Problem Formulation}
\label{sec:3_1}
To give formal definition of the problem, we separate it into two sub-tasks given the different representations of historical information. 

\noindent\textbf{Structured Event Forecasting (Graph\footnote{"Graph" is interchangeably used to represent this setting.})}. 
This formulation defines each event as a quadruple $(s, r, o, t)$, which is also called an atomic event, where $s, r, o, t$ corresponds to the subject, relation\footnote{Relation and event type are interchangeably used in this work}, object, and timestamp.
At each timestamp $t$, all the quadruples form an event graph, denoted as $G_t = \{(s, r, o, t)\}^N$, where $N$ is the number of events at timestamp $t$. Recent work\cite{LoGo} introduces the concept of complex event (CE) into the structured event representation by document clustering, elaborating the event’s occurrence situation or context. 
Specifically, each atomic event is extended from a quadruple to a quintuple, \ie $(s, r, o, t, c)$, where $s\in\mathcal{E}$, $r\in\mathcal{R}$, $o\in\mathcal{E}$, and $c\in\mathcal{C}$ represent the subject, relation, object, and CE, respectively; $\mathcal{E}$, $\mathcal{R}$ and $\mathcal{C}$ are the entity set, relation set and complex context set.
Correspondingly, the event graph at each timestamp will be extended as $G_t = \{(s, r, o, t, c)\}^N$. 
Furthermore, in addition event graph, there are images associated with structured events, denoted as $V_t = \{v_1, v_2, ..., v_m\}^M_{m=1}$, where $M$ is the number of images at timestamp $t$. Finally,
the structured event forecasting task can then be formulated as follows: given the historical event graphs $G_{<t}=\{G_0, G_1, ..., G_{t-1}\}$ and associated images $V_{<t}=\{V_0, V_1, ..., V_{t-1}\}$ before timestamp $t$, and a query $(s, r, t)$ or $(s, o, t)$, the goal is to predict the missing object $o$ or relation $r$. 

\noindent\textbf{Unstructured Event Forecasting (Text\footnote{"Text" is interchangeably used to represent this setting.})}. In addition to the structured event representation, we also consider the unstructured event representation, where the historical information is represented in the form of textual sub-events, \ie $A_t = \{a_1, a_2, ..., a_k\}^K_{k=1}$ and $A_t\in\mathcal{A}$, where $a_k$ denotes the k-th textual sub-events and $\mathcal{A}$ denotes the corpus of textual sub-events. The textual sub-events are obtained by summarizing the content of news articles. Similar to structured event forecasting, textual sub-events have associated images, denoted as $V_{<t}=\{V_0, V_1, ..., V_{t-1}\}$. The unstructured event forecasting task can be formulated as: given the historical textual sub-events $A_{<t}=\{A_0, A_1, ..., A_{t-1}\}$ and associated images $V_{<t}=\{V_0, V_1, ..., V_{t-1}\}$ before timestamp $t$, and a query $(s, r, t)$ or $(s, o, t)$, the goal is to predict the missing object $o$ or relation $s$.


\subsection{Image Function Identification}
\label{sec:3_2}
Identifying the function of images in the temporal event forecasting is the key to utilize multimodal visual information.
In news articles, images play a vital role not only in attracting readers but also in completing and enriching the textual content, especially key event content.
We will identify the image functions into three categories, \ie highlighting, complementary, and irrelevant, during the dataset construction stage.
Excluding the irrelevant images, the others serve distinct roles in the temporal event forecasting task. We propose an Image Function Identification module to recognize these functions as verbal descriptions using MLLMs, and subsequently incorporate these function descriptions into LLM-based forecasting models. Specifically, when the function of associated image is highlighting, the visual elements directly support and highlight the key sub-events described in the text. These "highlighting" sub-events, substantiated by corroborating information across modalities, can be identified as key events. To determine which sub-event is a key event, we leverage the MLLMs to analyze the images and sub-events from multiple aspects, including main objects, celebrities, activities, environment, and labeled items. 
In cases where the function of associated image is complementary, the visual content contains information that supplements and extends what is covered in the news text. 
To more effectively extract the relevant supplementary information, we consider the following aspects: 1) identify the main subject of the image as the central point, 2) directly relate the extracted information to the news event in the article, 3) prioritize the most newsworthy visual elements, 4) ensure all information comes directly from the provided news article without fabrication, and 5) aim for a concise summary using clear language. 
By analyzing the interplay between visual images and textual content within news articles, we can gain a more comprehensive understanding of the underlying events and better contextualize the temporal evolution of historical events. 
Ultimately, the prompts utilized in making predictions are shown below: 
\begin{verbatim}
SYSTEM: 
You are an assistant to perform event forecasting 
with the following rules:
1. The atomic event is the basic unit describing a spec-
ific event, typically presented in the form of a quadru-
ple (S, R, O, T), where S represents the subject, R repre-
sent the relation, O represents the object, and T repres-
ents the relative time.
2. When formulating the ultimate prediction, the preemi-
nent factor to be meticulously weighed and scrutinized 
is the [Key Events]. Complementing this paramount consi-
deration is the [Related events], which, though ancilla-
ry in nature, serves as a valuable adjunct, furnishing 
pertinent contextual details and auxiliary insights to 
fortify the predictive analysis.
3. Given a query of (S, R, T) in the future and the list 
of historical events until t, event forecasting aims to 
predict the missing object.
USER: 
[Query]: (S, O/R, T)
[Key Events]: xxx.
[Related Events]: xxx.
[Options]: A.xxx B.xxx C.xxx D.xxx E.xxx
\end{verbatim} 
The key events are explicitly highlighted within the prompts, while complementary information is provided as additional relevant events.

\subsection{Forecasting Framework}
\label{sec:3_3}
Given there are few established studies of using LLMs for event forecasting, we consider two representative approaches, \ie In-context Learning (ICL)~\cite{GPT-NeoX-ICL} and Retrieval Augmented Generation (RAG)~\cite{RAG}. Each of these two methods can accept both structured and unstructured historical input, and answer the structured forecasting questions.

\subsubsection{In-context Learning (ICL)}
In-context learning leverages both intrinsic and extrinsic factors to construct historical events. Specifically, the intrinsic factors of an event are related to its inherent elements, particularly the subject. In contrast, the extrinsic factors are driven by the contextual environment surrounding the event. Therefore, whether the data is structured or unstructured, we construct the historical events based on the subject and the complex event, separately. The details are as follows:
\begin{itemize}[leftmargin=*]
    \item \textbf{Structured Data.} For structured data, the method takes the discrete event graph as the input. To capture the intrinsic factors, we use the subject of the current event as a guiding clue to construct the historical event graph $\mathbf{G}_{<t}^{s}=\{G_0^s, G_1^s, ..., G_{t-1}^s\}$, where $G_t^s$ represents historical events graph at timestamp $t$ with the same subject as the current event. To account for the extrinsic factors, we construct the historical event graph from the complex event, \emph{i.e.} $\mathbf{G}_{<t}^{c}=\{G_0^c, G_1^c, ..., G_{t-1}^c\}$, where $G_t^c$ represents historical events graph at timestamp $t$ with the same complex event as the current event. Finally, with the highlighting and complementary functions of the images, the input historical event graph is $\mathbf{G}_{input}=[\textbf{G}_k, \textbf{G}_r, \textbf{G}_c]$, where $\mathbf{G}_{input}\in \mathbf{G}_{<t}^{s} \bigcup \mathbf{G}_{<t}^{c}$ and $\mathbf{G}_k$ denotes the key events, $\mathbf{G}_r$ represents the remaining events, and $\mathbf{G}_c$ corresponds to the complementary events, respectively.
    \item \textbf{Unstructured Data.} For unstructured data, the method takes the textual sub-events as input. Firstly, we identify the events by the historical events graph from the subject and complex event and find the corresponding textual sub-events set $\mathbf{A}_{<t}^{s}=\{A_0^s, A_1^s, ..., A_{t-1}^s\}$ and $\mathbf{A}_{<t}^{c}=\{A_0^c, A_1^c, ..., A_{t-1}^c\}$ through the relationships between textual sub-events and graph sub-events. Then, with the highlighting and complementary functions of the images, the input historical textual sub-events are similarly $\mathbf{A}_{input}=[\textbf{A}_k, \textbf{A}_r, \textbf{A}_c]$, where $\mathbf{A}_{input}\in \mathbf{A}_{<t}^{s} \bigcup \mathbf{A}_{<t}^{c}$ and $\mathbf{A}_k$ denotes the key events, $\mathbf{A}_r$ represents the remaining events, and $\mathbf{A}_c$ corresponds to the complementary events, respectively.
\end{itemize}

\subsubsection{Retrieval Augmented Generation (RAG)}
Despite the rich information provided by in-context learning methods, the inherent nature of the temporal event means that the existing historical event still contains substantial noise. Inspired by the recent research of RAG~\cite{RAG}, we also adopt the retrieve-then-generate paradigm to find the most relevant historical events to mitigate the problem of noise. Similar to ICL methods, we utilize two forms of data representation, structured data and unstructured data:
\begin{itemize}[leftmargin=*]
    \item \textbf{Structured Data.} Due to the structured nature of the data representation, the event graphs adhered to a unified quintuple format. Therefore, we first retrieve the entities that have interacted with the subject of the query event. Once we have obtained the related entity set, we can construct the history with the historical events where the subject or object is within this set. Similarly, through the function of images, the retrieval process also contains key events and complementary events.
    \item \textbf{Unstructured Data.} Unlike structured data, we can use the embedding techniques to directly retrieve relevant news events from a set of historical news articles for the unstructured data. Following this, we filter historical news events based on timestamps, eliminating outdated and irrelevant events. We also select the key events and complement information based on the images, which will be input according to the prompts described in Section \ref{sec:3_2}, and finally obtain the prediction results.
\end{itemize}

%% file: 4_experiments.tex
\section{Experiments}


\begin{table*}[htb]
\caption{Performance (accuracy) comparison between zero-shot LLM-based methods and the non-LLM methods in both settings of object entity prediction and relation prediction. For LLM-based methods, we include multiple backbones with two representative forecasting method, \ie ICL and RAG. 
Results of our methods are highlighted with grey backgrounds, where the key novelty lies in we leverage images by the Image Function Identification module.
}
\label{tbl:performance_compare}
\begin{tabular}{l|ll|cc|cc}
\toprule
\multirow{2}{*}{\textbf{Model Type/Backbone}}
& \multirow{2}{*}{\textbf{Forecasting Model}}
& \multirow{2}{*}{\textbf{Multimodal Model}} 
& \multicolumn{2}{c|}{Object Entity Prediction}        & \multicolumn{2}{c}{Relation Prediction}  \\  
&&&
\textbf{Text} & \textbf{Graph} & \textbf{Text} & \textbf{Graph}  \\ \midrule
\multirow{4}{*}{Non-LLM}  
& ConvTransE~\cite{ConvTransE} &  Uni-modal         
& N/A  & 0.3737  & N/A  & 0.7327  \\
& RGCN~\cite{RGCN}  &  Uni-modal       
& N/A  & 0.3777  & N/A  & 0.7203  \\ 
& RE-GCN~\cite{REGCN}  &  Uni-modal     
& N/A  & 0.3879  & N/A  & 0.7333  \\
& LoGo~\cite{LoGo} &  Uni-modal     
& N/A  & 0.3969  & N/A  & 0.7406  \\ \midrule

\multirow{2}{*}{Gemini-1.0-Pro-Vision\footnotemark[6]}
& \multirow{1}{*}{ICL~\cite{GPT-NeoX-ICL}}
& MLLM\footnotemark[6]  
& 0.3023  & 0.3319  & 0.5541  & 0.6085  \\
& \multirow{1}{*}{RAG~\cite{RAG}}
& MLLM\footnotemark[6]  
& 0.3305  & 0.3465  & 0.5769  & 0.5848  \\

\midrule

\multirow{4.4}{*}{Gemini-1.0-Pro\footnotemark[6]}

&\multirow{2}{*}{ICL~\cite{GPT-NeoX-ICL}}
& Uni-modal           
& 0.3312  & 0.3657  & 0.5900  & 0.6257  \\ 
&& \textbf{MM-Forecast (ours)}  \cellcolor[HTML]{F5F5F5}     
& \textbf{0.3527}  \cellcolor[HTML]{F5F5F5}
& \textbf{0.3837}  \cellcolor[HTML]{F5F5F5}
& \textbf{0.6087}  \cellcolor[HTML]{F5F5F5}
& \textbf{0.6324}  \cellcolor[HTML]{F5F5F5}  \\
\cmidrule{2-7}
&\multirow{2}{*}{RAG~\cite{RAG}}
& Uni-modal   
& 0.3340  & 0.3669  & 0.6081  & 0.5866  \\
&& \textbf{MM-Forecast (ours)}  \cellcolor[HTML]{F5F5F5}    
& \textbf{0.3425}  \cellcolor[HTML]{F5F5F5}  
& \textbf{0.3692}  \cellcolor[HTML]{F5F5F5}
& \textbf{0.6121}  \cellcolor[HTML]{F5F5F5}
& \textbf{0.5991}  \cellcolor[HTML]{F5F5F5}  \\ \midrule

\multirow{4.4}{*}{GPT-3.5-Turbo\footnotemark[7]}
& \multirow{2}{*}{ICL~\cite{GPT-NeoX-ICL}}
& Uni-modal        
& 0.3063  & 0.3431  & 0.4847  & 0.5345  \\   
&& \textbf{MM-Forecast (ours)}  \cellcolor[HTML]{F5F5F5} 
& \textbf{0.3414}  \cellcolor[HTML]{F5F5F5}
& \textbf{0.3522}  \cellcolor[HTML]{F5F5F5}
& \textbf{0.5317}  \cellcolor[HTML]{F5F5F5}
& \textbf{0.5521}  \cellcolor[HTML]{F5F5F5}  \\
\cmidrule{2-7}
& \multirow{2}{*}{RAG~\cite{RAG}}
& Uni-modal  
& 0.3272  & 0.3397  & 0.4943  & 0.4666  \\
&& \textbf{MM-Forecast (ours)}  \cellcolor[HTML]{F5F5F5} 
& \textbf{0.3652}  \cellcolor[HTML]{F5F5F5}  
& \textbf{0.3647}  \cellcolor[HTML]{F5F5F5}  
& \textbf{0.5152}  \cellcolor[HTML]{F5F5F5}
& \textbf{0.5113}  \cellcolor[HTML]{F5F5F5}  \\
\bottomrule
\end{tabular}
\end{table*}
We conduct experiments to evaluate the proposed approach, and answer the following research questions:

\begin{itemize}[leftmargin=*]
    \item \textbf{RQ1:} What is the overall performance of temporal event forecasting methods by including visual information?
    \item \textbf{RQ2:} How do the highlighting and complementary functions of images affect the forecasting performance?
    \item \textbf{RQ3:} How do different LLM backbones as well as fine-tuning affect the performance?
\end{itemize}

\subsection{Experimental Settings}

We introduce the experimental settings, including the dataset, the methods compared, and the implementation details.

\subsubsection{Dataset}
\label{sec:2_2}

We build our dataset based on MidEast-TE-mini~\cite{chang2024comprehensive}, which includes structured atomic events and news articles. We aim to add images that  correspond to the events in the dataset, hence we will have the data in visual modality. An intuitive way is to download the web page according to the URL provided by the original dataset. However, the original web page always contains a lot of irrelevant images, such as advertisement images, that are cumbersome and difficult to be accurately filtered out. Instead of directly solving this problem, we propose an alternative solution that we use Google Image Search\footnote{\url{https://images.google.com/}} to search the images using the news article title as the query.
Among the returned images, we select the top-ranked ones as the associated images of the news article. In order to further filter out irrelevant images, we instruct the Gemini-1.0-Pro-Vision model to determine the relevance of images to news articles. We give three options: highlighting, complementary and irrelevant. Highlighting means that the images and the content of the news are highly matched, and complementary means that the image has supplementary meaning to the content of the news. Images beyond these two are regarded as irrelevant. We further remove images that are classified as irrelevant.
Finally, we name our dataset as MidEast-TE-multimodal, short as MidEast-TE-mm.

\subsubsection{Compared Methods} 


The compared methods are categorised into non-LLM-based methods and LLM-based methods. For non-LLM-based methods, only text or graph modalities are involved, since these methods architecture are fixed. We train the models on the training set, selecting the best-performing model based on the validation set results, and obtain the final results of the testing set. For LLM-based methods, we use the proprietary LLMs due to their superior performance compared to open-source LLMs. Therefore, testing is generally done in a zero-shot manner, \ie directly test them on the testing set.
The specific methods are shown below:
\begin{itemize}[leftmargin=*]
    \item \textbf{ConvTransE~\cite{ConvTransE}:} This method employs a convolutional neural network (CNN) and a translational operation to capture the relational patterns within triplet data.
    \item \textbf{RGCN~\cite{RGCN}:} RGCN leverages a graph convolutional neural network (GCN) to capture the diverse relations between entities.
    \item \textbf{RE-GCN~\cite{REGCN}:} RE-GCN utilizes a combination of GCN and recurrent neural network (RNN) to capture both the relational patterns and temporal dynamics.
    \item \textbf{LoGo~\cite{LoGo}:} This method models relationships within and between complex events from both local and global perspectives.
    \item \textbf{GPT-3.5-Turbo\footnotemark[7]:} The GPT-3.5-turbo model is the prevalent iteration of the GPT (Generative Pre-trained Transformer) language model developed by OpenAI\footnote{\url{https://openai.com/}}. 
    \item \textbf{Gemini-1.0\footnotemark[6]:} Gemini-1.0 is a cutting-edge family of multimodal models developed by the Gemini Team at Google. 
\end{itemize}

\subsubsection{Implementation Details} 
To ensure the reproducibility, we fixed the temperature parameter of the proprietary LLMs used to 0 and set the seed parameter to a constant value. When making forecasting, we limit the maximum output token length to 256 to prevent invalid responses. To ensure fairness across the experiments, the history that can be retrieved is set to 30 days. Notably, the retrieval models that we employ include: BM25~\cite{BM25}, Contriever~\cite{Contriever}, and LlamaIndex~\cite{LlamaIndex}. Additionally, considering the limitation of the context window, we further restrict the maximum number of sub-events in the historical context to 50. Following previous methods~\cite{chang2024comprehensive}, we employ the Accuracy (Acc) as the evaluation metric. 
\footnotetext[6]{https://ai.google.dev/models/gemini}
\footnotetext[7]{https://platform.openai.com/docs/models/gpt-3-5-turbo}

\subsection{Performance Comparison (RQ1)}
We analyze our model's performance, by comparing various baseline methods on different experimental settings, different input forms, and different retrieval models.

\subsubsection{Performance \wrt Various Settings.}~\label{sec:4_3_1} The overall performance comparison is presented in the Table~\ref{tbl:performance_compare}. 
To comprehensively explore and evaluate methods, we conduct experiments across multiple dimensions, including the format of data representation (Text or Graph), the construction of historical information (RAG-based or ICL-based), and the prediction objective (Object or Relation). Clearly, we have the following observations.

First, enhancing LLM-based methods with visual information consistently improves their accuracy across all experimental settings.
This demonstrates that our proposed MM-Forecast makes effective use of visual information, leading to a better contextual understanding of historical information. Hence, our method strengthens the inference ability of LLM and makes more accurate event forecasting performance. 

Second, even though the performances of all LLM-based methods have been improved, they still under-perform the traditional Non-LLM based methods. 
The reason is that LLM-based methods are tested in zero-shot manner, while the Non-LLM methods, which follow supervised learning, are still competitive. Notably, by using our MM-Forecast method, LLM-based methods can achieve close or even better performance than Non-LLM methods for the object entity prediction task. 

\begin{table}[t!]
\caption{The results of using different retrieval models.}
\label{tbl:vary_retriever}
\begin{tabular}{lcc}
\toprule
\textbf{Retriever} & \textbf{Gemini-1.0-Pro} & \textbf{GPT-3.5-Turbo} \\
\midrule
\textbf{BM25}~\cite{BM25} & 0.3272 & 0.3318\\
\textbf{Contriver}~\cite{Contriever} & 0.3335 & 0.3431\\
\textbf{LlamaIndex}~\cite{LlamaIndex} & 0.3425 & 0.3652\\

\bottomrule
\end{tabular}
\vspace{-0.15in}
\end{table}

Third, the relation prediction task exhibits higher accuracy compared to the object entity prediction task. This suggests that the forecasting of entities is more challenging than relations. There are a few potential reasons for this. First, the set of entities (5909) is much larger than the set of relation types (267), so predicting specific entities is inherently more difficult given the larger candidate pool. 
Second, we deem that the information implied in entities is more explicit. Thus when two entities are given for a relation prediction, it is easier than when the subject and relation are given for an object prediction.

\begin{figure}[t]
\begin{center}
\includegraphics[width=\linewidth]{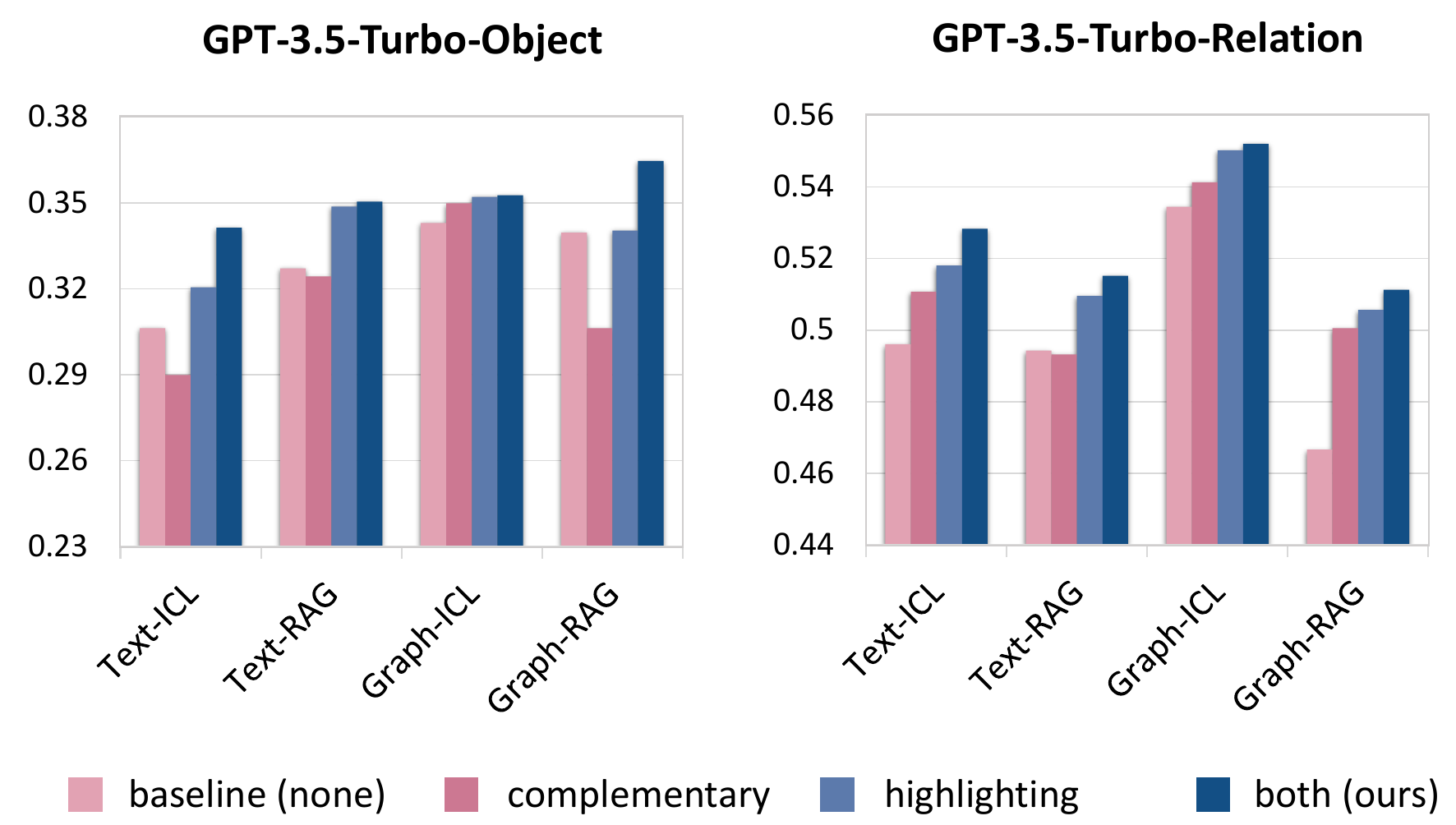}
\end{center}
\caption{Ablation study of each type of image functions.
}
\label{fig:ablation}
\vspace{-0.1in}
\end{figure}

\subsubsection{Performance \wrt Directly Using Images.} To illustrate the limitations of existing MLLMs in the task of temporal event forecasting, we also conduct experiments using the Gemini-1.0-Pro-Vsion model~\cite{gemini} and directly consuming the images in the sub-events. Specifically, this approach leverages the inherent image understanding capabilities of the Gemini-1.0-Pro-Vision model, which embeds image patches as features and seamlessly concatenates thes image features with textual features. From Table~\ref{tbl:performance_compare}, we can observe that the accuracy of using images directly is not only lower than our MM-Forecast, but also even worse than the method using only textual data (Uni-modal methods). 
This illustrates that existing proprietary MLLMs still struggle to make effective event forecasting with multiple images, and reflects the superiority of our MM-Forecast.

\subsubsection{Performance \wrt Various Retrieval Models.} The choice of retrieval model may have a significant impact on forecasting. 
The experiments here involve only unstructured event forecasting, since the structured approach employs retrieval based on keyword search techniques.
To explore the effect of retrieval model, we adopt three different retrieval models, \ie BM25~\cite{BM25}, Contriver~\cite{Contriever}, and LLamaIndex~\cite{LlamaIndex}, then equip them into our forecasting framework, and obtain the forecasting results.
From the results in Table~\ref{tbl:vary_retriever}, we can observe that the performance progressively improves by using stronger retrieval models, with LLamaIndex performing the best, followed by Contriver, and then BM25. These results verify that stronger retrieval capabilities lead to better forecasting performance, suggesting that retrieval-oriented method design is a promising direction for future research. 
This phenomenon is consistent with the observation concluded from recent works~\cite{chang2024comprehensive}.

\subsection{Study of the Image Functions (RQ2)}
\subsubsection{Effects of Image Functions} We conduct ablation experiments for the highlighting and complementary function of images. The results are shown in Figure~\ref{fig:ablation}. First, the model that leverages both the highlighting of key events and the complementary information performs the best across the experimental settings. In addition, the performance of the model with only key events highlighted is sub-optimal. 
This illustrates the effectiveness of the highlighting function of images, and it elicit the fact that highlighting and complementary reinforce each other to achieve even better prediction results. 
Second, we can observe that in some settings (Text-ICL, Text-RAG), the performances of the model with only complementary information are even worse than the baseline model. The possible reason for this is that the offering of complementary information also introduces more noise and therefore leads the degradation of performance. 
Third, the performance of RAG-based method is obviously worse than the ICL-based method in the relation prediction task, meanwhile, such performance gap does not exist in the entity prediction task. This is may because that relation prediction is easier than object entity prediction, as mentioned in section~\ref{sec:4_3_1}. As a result, ICL-based historical events may already contain enough information to make accurate relation prediction, whereas the retrieval model may not retrieve relevant information instead. 

\begin{table}[t!]
\caption{The accuracy of image function identification.}
\label{tbl:rel_acc}
\begin{tabular}{lcccc}
\toprule

\multirow{2}{*}{\textbf{Data-Type}} 
& \multicolumn{2}{c}{\textbf{GPT-4-Vision}} &\multicolumn{2}{c}{\textbf{Human}}\\
& Text & Graph & Text & Graph \\
\midrule
\textbf{Highlighting} & 0.68 & 0.68 & 0.73 & 0.83\\
\textbf{Complementary} & 0.88 & 0.93 & 0.87 & 0.86\\

\bottomrule
\end{tabular}
\end{table}




\begin{table}
\caption{Result comparison between using our identified and randomly-assigned image functions.}
\label{tbl:random_rel}
\begin{tabular}{ll|cc|cc}
\toprule
\multirow{2}{*}{\textbf{Model}}              
& \multirow{2}{*}{\textbf{Settings}} 
& \multicolumn{2}{c|}{Object}        & \multicolumn{2}{c}{Relation}  \\  
&&
\textbf{Text} & \textbf{Graph} & \textbf{Text} & \textbf{Graph}  \\ \midrule
\multirow{2}{*}{GPT-3.5-Turbo}    
& \textbf{Random}  
& 0.3284 & 0.3394  & 0.5156  & 0.5249  \\ 
& \textbf{Ours}         
& 0.3414  & 0.3522  & 0.5317  & 0.5521  \\    
\bottomrule
\end{tabular}
\vspace{-0.05in}
\end{table}

\begin{figure*}
\begin{center}
\includegraphics[width=\linewidth]{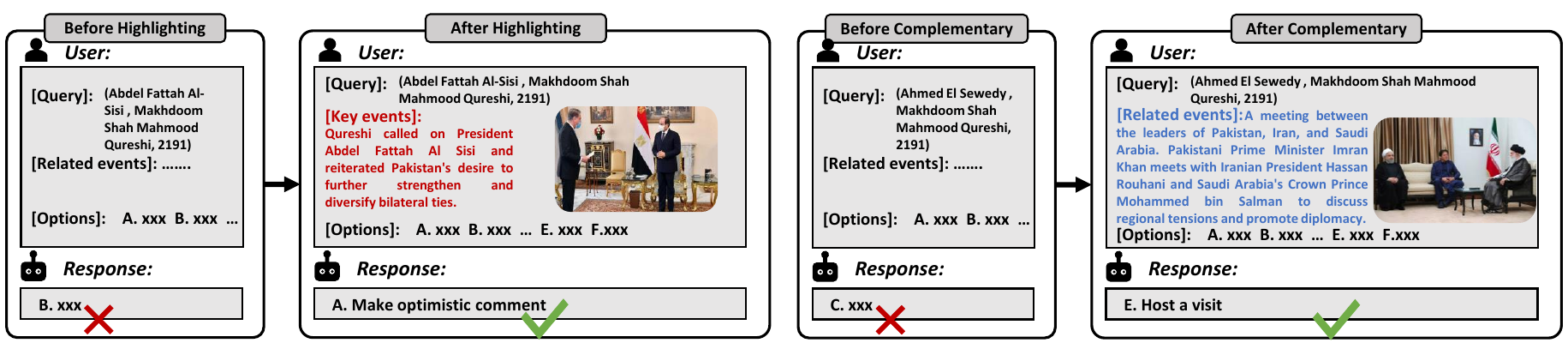}
\end{center}
\caption{Case study: two examples that when considering \textit{highlighting} and \textit{complementary} functions of images, our method yields better forecasting results compared with the baselines.
}
\label{fig:casestudy}
\vspace{-0.1in}
\end{figure*}

\subsubsection{Analysis of the Image Function Identification} 
In addition to overall forecasting performance analysis, we conduct in-depth study to directly assess the efficacy of highlighting and complementary function.
Specifically, we design additional experiments at the data level and prompt level to further verify the function of images. 
At the data level, we randomly sample 100 images of two categories respectively, and then judge the correctness of the classification by the powerful MLLM GPT-4-Vision~\footnote[8]{https://platform.openai.com/docs/models/gpt-4-turbo-and-gpt-4}.
As shown in Table~\ref{tbl:rel_acc}, both classification of highlighting and complementary functions show high accuracy. Furthermore, we can observe that the accuracy of highlighting is lower than that of complementary on all settings, which should be due to its more strict definition. The high accuracy of image functions in both LLM and human identification indicate that the images we used can indeed play the highlighting and complementary functions. 
In addition to direct assessment of the quality of image function identification, we conduct another ablation study by replacing our identified functions with randomly selected sub-events. Looking into the forecasting results in Table~\ref{tbl:random_rel}, random selection of sub-events leads to a decrease in forecasting accuracy, indicating that correct image function identification is crucial to the forecasting.

Finally, on top of quantitative evaluation, we conduct qualitative analysis and demonstrate two examples in Figure~\ref{fig:casestudy}. The first image emphasizes the event of Makhdoom Shah Mahmood Qureshi's visit to Abdel Fattah Al-Sisi, highlighting their efforts to strengthen and diversify bilateral relations. This highlighting function leads to a correct prediction of the event type. The second image provides supplementary information about the meeting between the two politicians, enabling an accurate prediction of the question. 

\subsection{Performance on Open-source and Fine-tuned LLMs (RQ3)}
All of the above LLM-based forecasting backbones  are implemented using proprietary LLMs in the zero-shot manner without any finetuning. We are interested in how our method performs on open-source LLMs, especially finetuned open-source LLMs.
Addressing this intriguing question, we select one of the most popular open-source LLMs, \ie Vicuna-7b, to replace the forecasting backbone LLM in our framework, with both zero-shot manner and fine-tune following typical instruction tuning with QLoRA~\cite{QLoRA}.
The results of object entity prediction are presented in Table~\ref{tbl:finetune-LLM}, which also includes the best results for proprietary LLMs and non-LLM methods. We observe that the zero-shot performance of Vicuna-7B is worse than its corresponding performance on proprietary LLMs, owing to the inherent capacity gap. However, after fine-tuning, Vicuna-7B achieves substantial performance gains, not only surpassing the proprietary LLMs but also outperforming all the non-LLM methods. 
In addition to fine-tuning the LLMs on object entity prediction, we also fine-tuning on the relation prediction task, as shown in Table~\ref{tbl:finetune-LLM-relation}. In both the text and the graph settings, the relation prediction results are consistent with the entity prediction, \ie fine-tuned LLMs achieve the best performance.
These results demonstrate the significant potential of fine-tuning LLMs for the temporal event forecasting task. 

\begin{table}
\caption{Performance of fine-tuned LLMs and its comparison with proprietary LLMs and non-LLM methods.}
\label{tbl:finetune-LLM}
\resizebox{\linewidth}{!}{
\begin{tabular}{clccc}
\toprule

\multicolumn{2}{c}{\textbf{Model}} 
& \textbf{Vicuna-7b} & \textbf{LLM} & \textbf{Non-LLM} \\
\midrule
\multirow{2}{*}{zero-shot} 
& \textbf{MM-Forecast-text-h} 
& 0.2723 & 0.3527 & N/A  \\
& \textbf{MM-Forecast-graph-h} 
& 0.2502 & 0.3837 & N/A  \\
\midrule
\multirow{2}{*}{fine-tune} 
& \textbf{MM-Forecast-text-h} 
& 0.4490 & N/A & N/A  \\
& \textbf{MM-Forecast-graph-h} 
& 0.5480 & N/A & 0.3969  \\

\bottomrule
\end{tabular}
}
\vspace{-0.1in}
\end{table}

\begin{table}
\caption{Performance of FT LLMs on the relation prediction.}
\label{tbl:finetune-LLM-relation}
\resizebox{\linewidth}{!}{
\begin{tabular}{lccc}
\toprule

\multicolumn{1}{c}{\textbf{Model}} 
& \textbf{Vicuna-7b} & \textbf{LLM} & \textbf{Non-LLM} \\
\midrule
\textbf{MM-Forecast-text-h} 
& \textbf{0.7809} & 0.6087 & N/A  \\
\textbf{MM-Forecast-graph-h} 
& \textbf{0.7901} & 0.6324 & 0.7406  \\

\bottomrule
\end{tabular}
}
\end{table}

%% file: 5_conclusion.tex
\section{Conclusion and Future Work}
In this paper, we studied an emerging and interesting problem of multimodal temporal event forecasting. We identified two essential image functions in the scenario of temporal event forecasting, \ie highlighting and complementary.
Then, we introduced MM-Forecast, a novel framework that leverages visual information to enhance temporal event forecasting. By recognizing the highlighting and complementary functions of images and translating them into verbal descriptions, we were able to seamlessly integrate this visual information into LLM-based forecasting models. Ultimately, this enabled the integration of visual information to enhance temporal event forecasting task. 


Looking ahead, there are numerous avenues for future work to address the key challenges. In particular, we would like to highlight three distinct aspects that warrant further exploration. First, multi-images relationship need to be considered. There are inherent relationships between images in related historical events, and these relationships are also important for event forecasting. Second, seeing is believing. Images have significant effects on the event forecasting task rather than accuracy improvement, that is credibility or trustability. Third, our current solution is still a multi-step pipeline, while devising an end-to-end approach using MLLMs is intriguing to explore in the future. 

%% file: 6_ack.tex
\begin{acks}
This work is partially supported by the National Natural Science Foundation of China under grant 62220106008, U20B2063 and 62102070. This work is also partially supported by Sichuan Science and Technology Program under grant 2023NSFSC1392. This research is also supported by Asian Institute of Digital Finance and NExT Research Center.
\end{acks}

%% file: 7_appendix.tex
\section{Appendix}

\begin{table*}[h]
\caption{Prompts of image function identification module.}
\label{tbl:supply_prompt}
\begin{tabular}{l|l}
\toprule
\multirow{10}{*}{\textbf{Identification}} & You are a professional news writer. \\
& Please judge the relationship between images and news based on the following rules: \\
& 1. Final judgment please choose between [highlighting, complementary, irrelevant].\\
& 2. The relationship between an image and a news article is highlighting if the image's subject matter and depicted \\
& \ \ \ \ event are highly related to the news and the specific event shown in the image is already mentioned in detail in\\
& \ \ \ \ the article's description.\\
& 3. The relationship between an image and a news article is complementary if the image's overall theme and \\
& \ \ \ \ background information are highly related to the news, but the specific event depicted in the image is not \\
& \ \ \ \ mentioned in detail in the article, and the visual information in the image can complement the news story as \\
& \ \ \ \ a whole.\\
&4. Except in cases where the relationship is highlighting or complementary, in other cases, the relationship \\
& \ \ \ \ between the image and the text is irrelevant. \\

\midrule
\multirow{14}{*}{\textbf{Highlighting}} & You are a professional news writer. \\
& Please determine which sub-event in the news the image is most relevant to based on the following rules:\\
& 1. For the final judgement, please answer with the serial number of the sub-event. For example: [The number of \\
& \ \ \ \ the sub-event most relevant to the image is 1.] \\
& 2. Identify the main subjects or objects prominently featured in the image. Sub-events that provide details, \\
& \ \ \ \ background information or context directly about these central visual elements are highly relevant. \\
&3. If people are depicted, identify who those individuals are. Sub-events involving those particular people should \\
& \ \ \ \ take priority. \\
&4. Analyze the overall activities, actions, emotions or mood being portrayed in the image. Relevant sub-events \\
& \ \ \ \ likely delve into similar situations, occurrences or sentiments illustrated. \\
&5. Take note of the specific location, setting or environment depicted in the image. Prioritize sub-events that \\
& \ \ \ \ discuss that geographic area, type of place, or related events. \\
&6. Look for any text, logos, labeled items or signs visible in the image content. Sub-events elaborating on the \\
& \ \ \ \ organizations, companies, products or public figures represented by those texts are applicable.\\

\midrule
\multirow{18}{*}{\textbf{Complementary}} & You are a professional news writer.\\
& Please extract the image information according to the following rules based on the content of the provided news: \\
& 1. Extract the image information as a sub-event. Instead of multiple sub-events.\\
& 2. The phrases: [In the image], [The image shows], [In the picture], [The image is], [In the photo], etc, should \\
& \ \ \ \ never appear in the summarised sub-event.\\
& 3. Identify the primary focus or subject of the image that represents the core piece of information being conveyed. \\
& \ \ \ \ This main subject should serve as the central point around which the image information is extracted.\\
& 4. Directly relate the extracted image information to the associated news event covered in the article. The image \\
& \ \ \ \ summary should complement and enhance the understanding of the news content, not introduce unrelated \\
& \ \ \ \ information.\\
& 5. Prioritize and emphasize the most newsworthy and significant details visible in the image. These could include \\
& \ \ \ \ specific actions, emotions, or identifying characteristics of the main subject. \\
& 6. Ensure that all information included in the image summary originates directly from the provided image \\
& \ \ \ \ and news article. Avoid introducing fabricated content, speculative details. \\
& 7. Aim for a succinct summary, using clear and straightforward language. Avoid excessive detail or subjective \\
& \ \ \ \ commentary. \\
& 8. Maintain an objective and impartial tone when describing the image. Avoid inserting personal opinions or \\
& \ \ \ \ interpretations. \\
\bottomrule
\end{tabular}
\end{table*}

\subsection{Prompts: Image Function}
In this section, we show all the prompts that need to be used in the image function identification module. As show in Table~\ref{tbl:supply_prompt}, the first row is the prompt for image function recognition, which is mainly from the perspective of the subject background and the specific event to judge the function of the image. The last two rows are the prompts of the different functions of the images to achieve their respective functions and transform their information into verbal descriptions. Eventually, the verbal information will be integrated into the LLM-based event forecasting model.

\subsection{Case study: Image Function}
To further illustrate that our approach does indeed identify truly key events and the required complementary information, we provide additional examples. In the first example of the highlighting function, the image directly depicts Ocasio-Cortez, with the background appearing to be the Congressional sites, thereby emphasizing the relevant key event. Correspondingly, the key event also mentions the relationship between Congress and Ocasio-Cortez. Consequently, an accurate prediction is achieved.
In the second example of the highlighting function, the key event highlighted by the image directly mentions the disqualification of Ali Larijani from the election, which perfectly aligns with the results that need to be predicted and the information provided to present those results.
For the first example of complementary functions, the image provides information about the signing of a free trade agreement between Turkey and the United Kingdom. While enhanced trade has the potential to lead to employment and economic growth, the image offers complementary information on the role of labor. Therefore, an accurate forecast is achieved.
In the second example about the complementary function, the image shows Bernie Sanders who is a democratic progressive socialist like Ocasio-Cortez. They share many commonalities and connections to Congress, which can provide supplementary information to more accurately predict the outcome.
Through these examples, the distinct functions of highlighting key events and providing complementary information are elucidated, substantiating the effectiveness of our approach in leveraging multimodal information for accurate temporal event forecasting.



\begin{figure*}
\centering
\includegraphics[width=0.98\textwidth, height=\textheight]{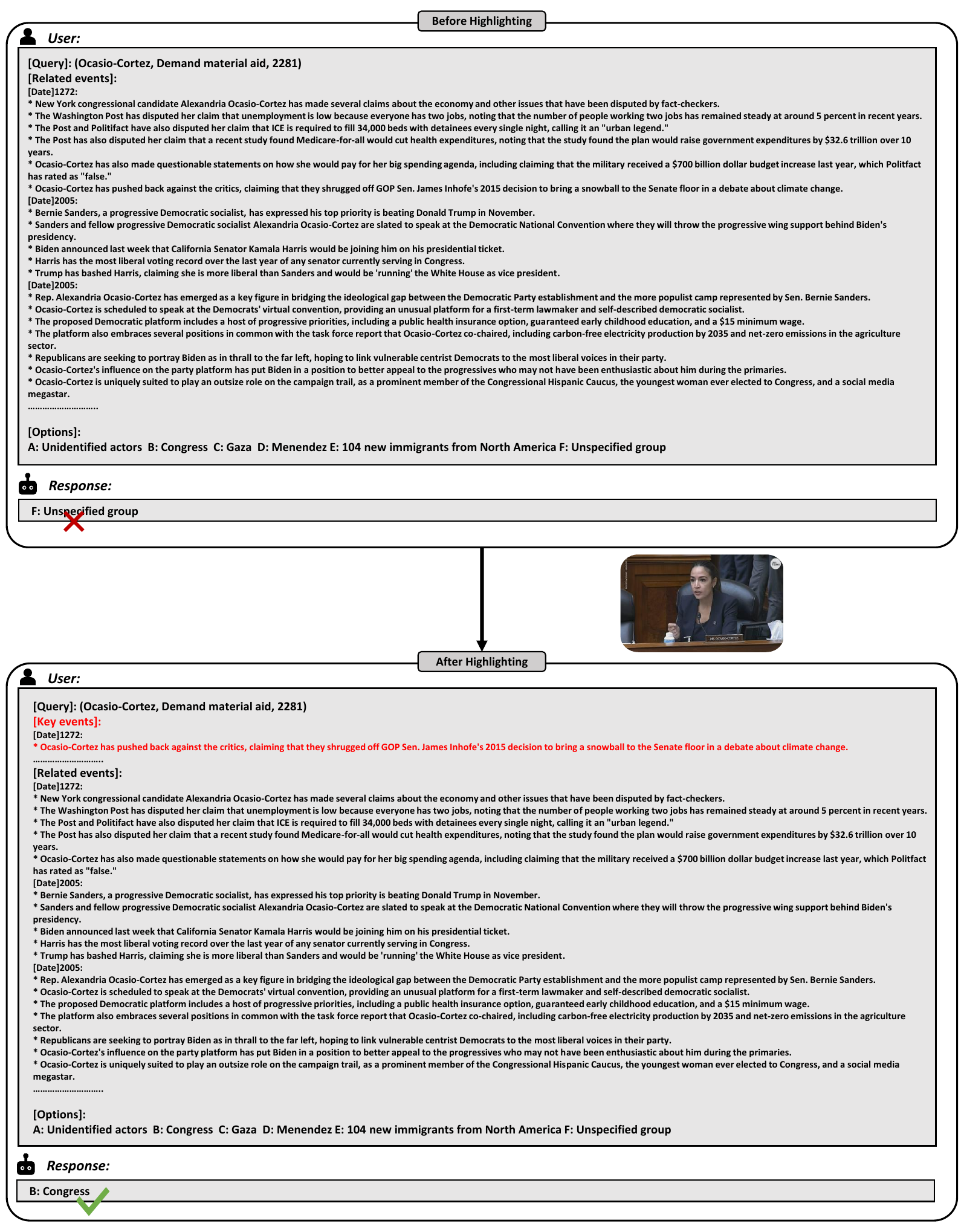}

\caption{The case study of \textit{highlighting} function of image.
}
\label{fig:supply_case_study}
\end{figure*}

\begin{figure*}
\centering
\includegraphics[width=0.98\textwidth, height=\textheight]{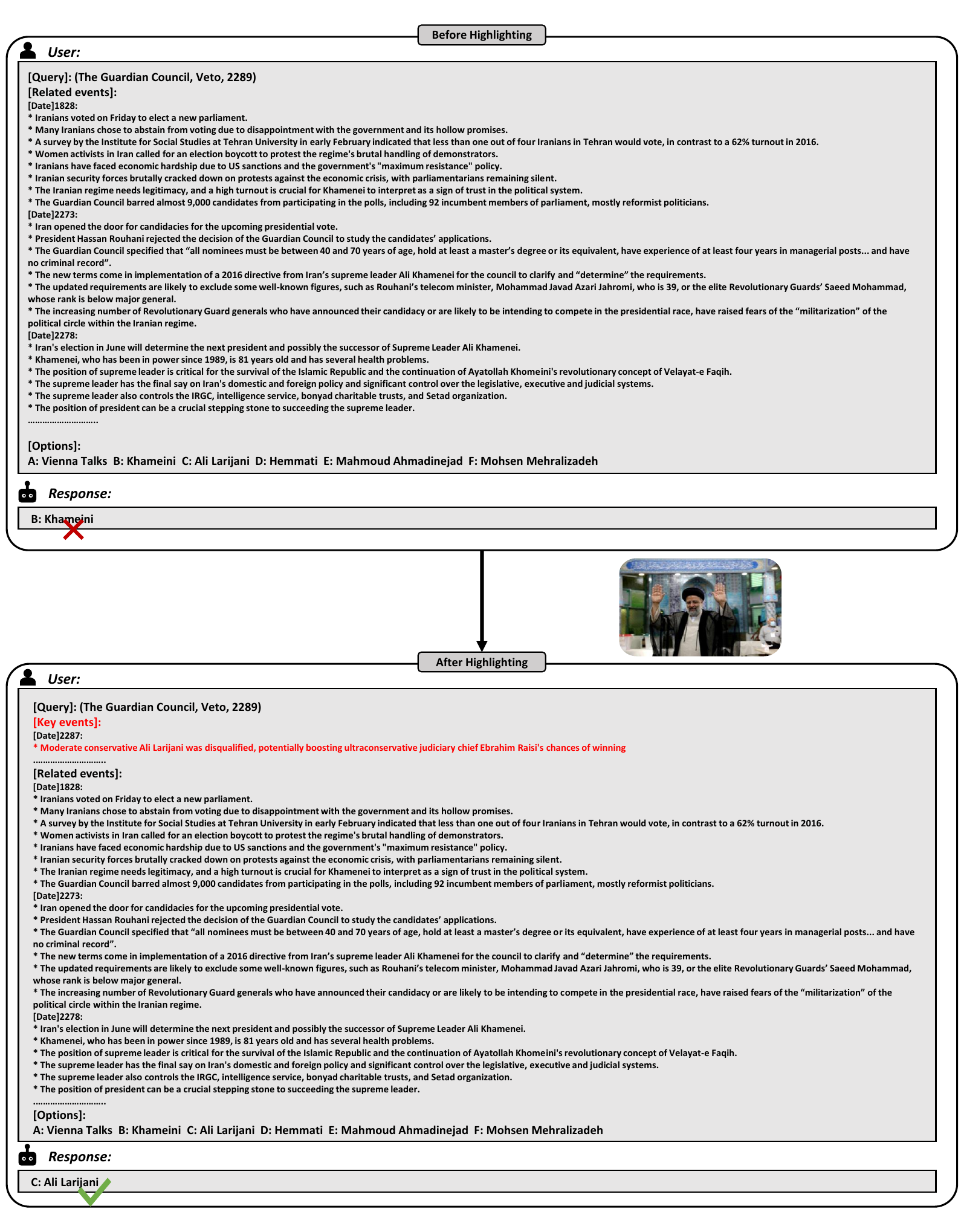}

\caption{The case study of \textit{highlighting} function of image.
}
\label{fig:supply_case_study}
\end{figure*}

\begin{figure*}
\centering
\includegraphics[width=0.98\textwidth, height=\textheight]{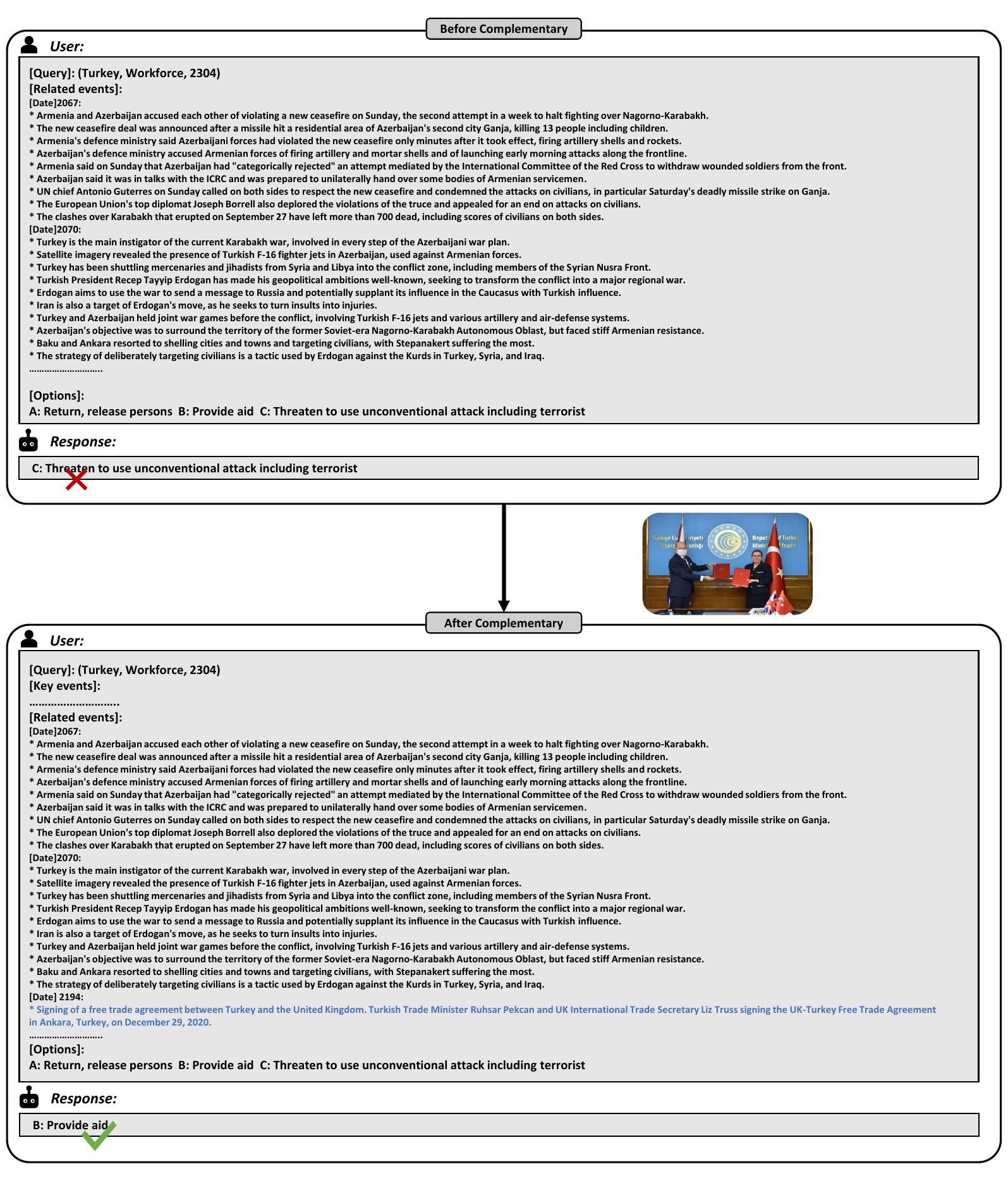}

\caption{The case study of \textit{complementary} function of image.
}
\label{fig:supply_case_study}
\end{figure*}

\begin{figure*}
\centering
\includegraphics[width=0.98\textwidth, height=\textheight]{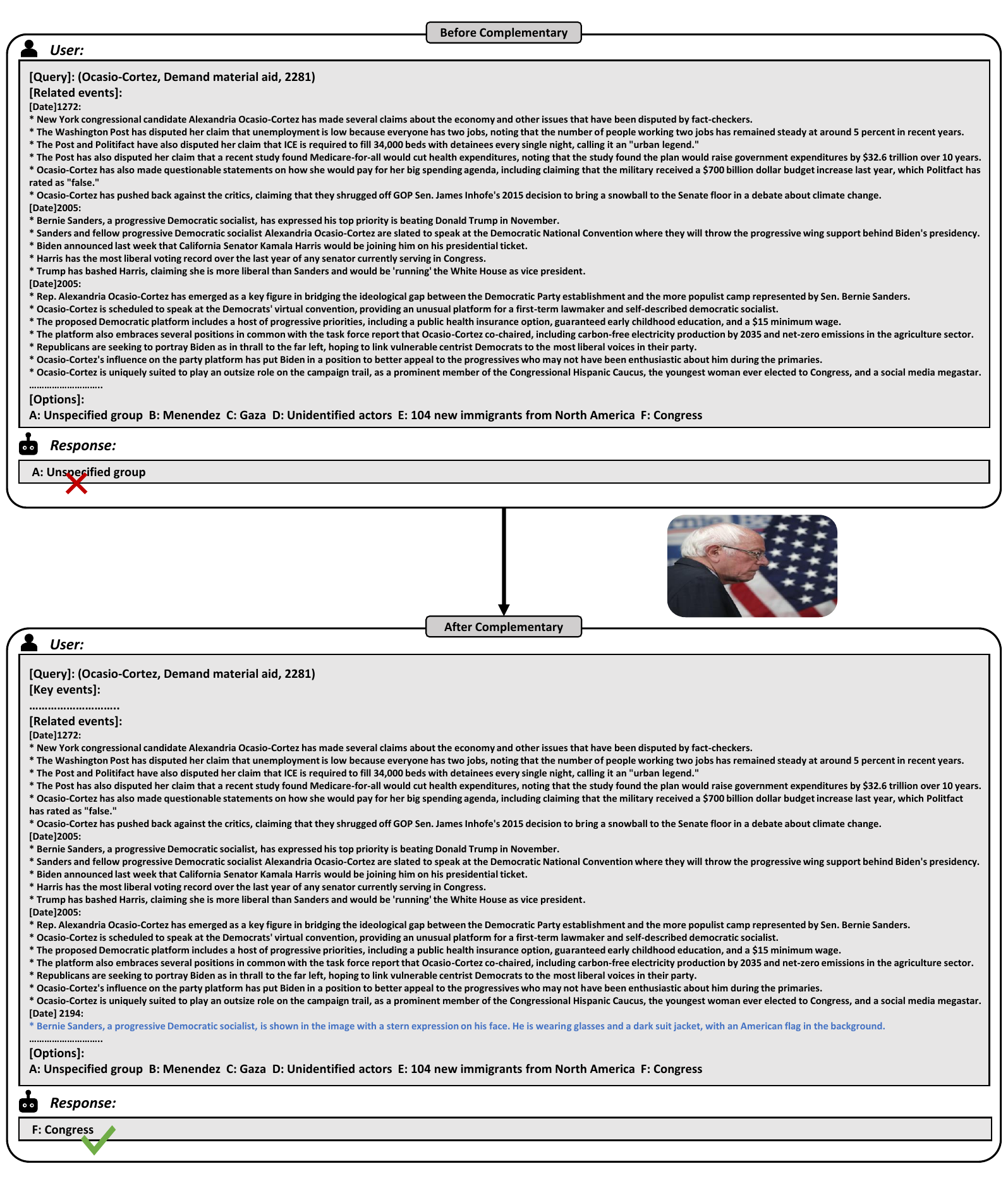}

\caption{The case study of \textit{complementary} function of image.
}
\label{fig:supply_case_study}
\end{figure*}

%% file: 0_main.bbl

\begin{thebibliography}{51}


\ifx \showCODEN    \undefined \def \showCODEN     #1{\unskip}     \fi
\ifx \showDOI      \undefined \def \showDOI       #1{#1}\fi
\ifx \showISBNx    \undefined \def \showISBNx     #1{\unskip}     \fi
\ifx \showISBNxiii \undefined \def \showISBNxiii  #1{\unskip}     \fi
\ifx \showISSN     \undefined \def \showISSN      #1{\unskip}     \fi
\ifx \showLCCN     \undefined \def \showLCCN      #1{\unskip}     \fi
\ifx \shownote     \undefined \def \shownote      #1{#1}          \fi
\ifx \showarticletitle \undefined \def \showarticletitle #1{#1}   \fi
\ifx \showURL      \undefined \def \showURL       {\relax}        \fi
\providecommand\bibfield[2]{#2}
\providecommand\bibinfo[2]{#2}
\providecommand\natexlab[1]{#1}
\providecommand\showeprint[2][]{arXiv:#2}

\bibitem[Alayrac et~al\mbox{.}(2022)]%
        {Flamingo}
\bibfield{author}{\bibinfo{person}{Jean-Baptiste Alayrac}, \bibinfo{person}{Jeff Donahue}, \bibinfo{person}{Pauline Luc}, \bibinfo{person}{Antoine Miech}, \bibinfo{person}{Iain Barr}, \bibinfo{person}{Yana Hasson}, \bibinfo{person}{Karel Lenc}, \bibinfo{person}{Arthur Mensch}, \bibinfo{person}{Katherine Millican}, \bibinfo{person}{Malcolm Reynolds}, {et~al\mbox{.}}} \bibinfo{year}{2022}\natexlab{}.
\newblock \showarticletitle{Flamingo: a visual language model for few-shot learning}. In \bibinfo{booktitle}{\emph{NeurIPS}}.
\newblock


\bibitem[Benjamin et~al\mbox{.}(2023)]%
        {benjamin2023hybrid}
\bibfield{author}{\bibinfo{person}{Daniel~M Benjamin}, \bibinfo{person}{Fred Morstatter}, \bibinfo{person}{Ali~E Abbas}, \bibinfo{person}{Andres Abeliuk}, \bibinfo{person}{Pavel Atanasov}, \bibinfo{person}{Stephen Bennett}, \bibinfo{person}{Andreas Beger}, \bibinfo{person}{Saurabh Birari}, \bibinfo{person}{David~V Budescu}, \bibinfo{person}{Michele Catasta}, {et~al\mbox{.}}} \bibinfo{year}{2023}\natexlab{}.
\newblock \showarticletitle{Hybrid forecasting of geopolitical events}.
\newblock \bibinfo{journal}{\emph{AI Magazine}} (\bibinfo{year}{2023}).
\newblock


\bibitem[Bin et~al\mbox{.}(2023)]%
        {bin2023unifying}
\bibfield{author}{\bibinfo{person}{Yi Bin}, \bibinfo{person}{Haoxuan Li}, \bibinfo{person}{Yahui Xu}, \bibinfo{person}{Xing Xu}, \bibinfo{person}{Yang Yang}, {and} \bibinfo{person}{Heng~Tao Shen}.} \bibinfo{year}{2023}\natexlab{}.
\newblock \showarticletitle{Unifying two-stream encoders with transformers for cross-modal retrieval}. In \bibinfo{booktitle}{\emph{Proceedings of the 31st ACM International Conference on Multimedia}}. \bibinfo{pages}{3041--3050}.
\newblock


\bibitem[Chang et~al\mbox{.}(2024)]%
        {chang2024comprehensive}
\bibfield{author}{\bibinfo{person}{He Chang}, \bibinfo{person}{Chenchen Ye}, \bibinfo{person}{Zhulin Tao}, \bibinfo{person}{Jie Wu}, \bibinfo{person}{Zhengmao Yang}, \bibinfo{person}{Yunshan Ma}, \bibinfo{person}{Xianglin Huang}, {and} \bibinfo{person}{Tat-Seng Chua}.} \bibinfo{year}{2024}\natexlab{}.
\newblock \showarticletitle{A Comprehensive Evaluation of Large Language Models on Temporal Event Forecasting}.
\newblock \bibinfo{journal}{\emph{arXiv preprint arXiv:2407.11638}} (\bibinfo{year}{2024}).
\newblock


\bibitem[Chiang et~al\mbox{.}(2023)]%
        {vicuna2023}
\bibfield{author}{\bibinfo{person}{Wei-Lin Chiang}, \bibinfo{person}{Zhuohan Li}, \bibinfo{person}{Zi Lin}, \bibinfo{person}{Ying Sheng}, \bibinfo{person}{Zhanghao Wu}, \bibinfo{person}{Hao Zhang}, \bibinfo{person}{Lianmin Zheng}, \bibinfo{person}{Siyuan Zhuang}, \bibinfo{person}{Yonghao Zhuang}, \bibinfo{person}{Joseph~E. Gonzalez}, \bibinfo{person}{Ion Stoica}, {and} \bibinfo{person}{Eric~P. Xing}.} \bibinfo{year}{2023}\natexlab{}.
\newblock \bibinfo{title}{Vicuna: An Open-Source Chatbot Impressing GPT-4 with 90\%* ChatGPT Quality}.
\newblock
\newblock
\urldef\tempurl%
\url{https://lmsys.org/blog/2023-03-30-vicuna/}
\showURL{%
\tempurl}


\bibitem[Chowdhery et~al\mbox{.}(2023)]%
        {palm}
\bibfield{author}{\bibinfo{person}{Aakanksha Chowdhery}, \bibinfo{person}{Sharan Narang}, \bibinfo{person}{Jacob Devlin}, \bibinfo{person}{Maarten Bosma}, \bibinfo{person}{Gaurav Mishra}, \bibinfo{person}{Adam Roberts}, \bibinfo{person}{Paul Barham}, \bibinfo{person}{Hyung~Won Chung}, \bibinfo{person}{Charles Sutton}, \bibinfo{person}{Sebastian Gehrmann}, {et~al\mbox{.}}} \bibinfo{year}{2023}\natexlab{}.
\newblock \showarticletitle{Palm: Scaling language modeling with pathways}.
\newblock \bibinfo{journal}{\emph{Journal of Machine Learning Research}} \bibinfo{volume}{24}, \bibinfo{number}{240} (\bibinfo{year}{2023}), \bibinfo{pages}{1--113}.
\newblock


\bibitem[Deng et~al\mbox{.}(2024)]%
        {deng2024advances}
\bibfield{author}{\bibinfo{person}{Songgaojun Deng}, \bibinfo{person}{Maarten de Rijke}, {and} \bibinfo{person}{Yue Ning}.} \bibinfo{year}{2024}\natexlab{}.
\newblock \showarticletitle{Advances in Human Event Modeling: From Graph Neural Networks to Language Models}.
\newblock  (\bibinfo{year}{2024}).
\newblock


\bibitem[Dettmers et~al\mbox{.}(2018)]%
        {ConvE}
\bibfield{author}{\bibinfo{person}{Tim Dettmers}, \bibinfo{person}{Pasquale Minervini}, \bibinfo{person}{Pontus Stenetorp}, {and} \bibinfo{person}{Sebastian Riedel}.} \bibinfo{year}{2018}\natexlab{}.
\newblock \showarticletitle{Convolutional 2D Knowledge Graph Embeddings}. In \bibinfo{booktitle}{\emph{{AAAI}}}. \bibinfo{publisher}{{AAAI} Press}, \bibinfo{pages}{1811--1818}.
\newblock


\bibitem[Dettmers et~al\mbox{.}(2023)]%
        {QLoRA}
\bibfield{author}{\bibinfo{person}{Tim Dettmers}, \bibinfo{person}{Artidoro Pagnoni}, \bibinfo{person}{Ari Holtzman}, {and} \bibinfo{person}{Luke Zettlemoyer}.} \bibinfo{year}{2023}\natexlab{}.
\newblock \showarticletitle{QLoRA: Efficient Finetuning of Quantized LLMs}.
\newblock \bibinfo{journal}{\emph{CoRR}}  \bibinfo{volume}{abs/2305.14314} (\bibinfo{year}{2023}).
\newblock


\bibitem[Ding et~al\mbox{.}(2024)]%
        {ding2024fashionregen}
\bibfield{author}{\bibinfo{person}{Yujuan Ding}, \bibinfo{person}{Yunshan Ma}, \bibinfo{person}{Wenqi Fan}, \bibinfo{person}{Yige Yao}, \bibinfo{person}{Tat-Seng Chua}, {and} \bibinfo{person}{Qing Li}.} \bibinfo{year}{2024}\natexlab{}.
\newblock \showarticletitle{Fashionregen: Llm-empowered fashion report generation}. In \bibinfo{booktitle}{\emph{Companion Proceedings of the ACM on Web Conference 2024}}. \bibinfo{pages}{991--994}.
\newblock


\bibitem[Gholipour~Ghalandari et~al\mbox{.}(2020)]%
        {gholipour-ghalandari-etal-2020-large}
\bibfield{author}{\bibinfo{person}{Demian Gholipour~Ghalandari}, \bibinfo{person}{Chris Hokamp}, \bibinfo{person}{Nghia~The Pham}, \bibinfo{person}{John Glover}, {and} \bibinfo{person}{Georgiana Ifrim}.} \bibinfo{year}{2020}\natexlab{}.
\newblock \showarticletitle{A Large-Scale Multi-Document Summarization Dataset from the {W}ikipedia Current Events Portal}. In \bibinfo{booktitle}{\emph{Proceedings of the 58th Annual Meeting of the Association for Computational Linguistics}}, \bibfield{editor}{\bibinfo{person}{Dan Jurafsky}, \bibinfo{person}{Joyce Chai}, \bibinfo{person}{Natalie Schluter}, {and} \bibinfo{person}{Joel Tetreault}} (Eds.). \bibinfo{publisher}{Association for Computational Linguistics}, \bibinfo{address}{Online}, \bibinfo{pages}{1302--1308}.
\newblock
\urldef\tempurl%
\url{https://doi.org/10.18653/v1/2020.acl-main.120}
\showDOI{\tempurl}


\bibitem[Izacard et~al\mbox{.}(2021)]%
        {Contriever}
\bibfield{author}{\bibinfo{person}{Gautier Izacard}, \bibinfo{person}{Mathilde Caron}, \bibinfo{person}{Lucas Hosseini}, \bibinfo{person}{Sebastian Riedel}, \bibinfo{person}{Piotr Bojanowski}, \bibinfo{person}{Armand Joulin}, {and} \bibinfo{person}{Edouard Grave}.} \bibinfo{year}{2021}\natexlab{}.
\newblock \showarticletitle{Unsupervised dense information retrieval with contrastive learning}.
\newblock \bibinfo{journal}{\emph{arXiv preprint arXiv:2112.09118}} (\bibinfo{year}{2021}).
\newblock


\bibitem[Jiao et~al\mbox{.}(2023)]%
        {eventChain}
\bibfield{author}{\bibinfo{person}{Yizhu Jiao}, \bibinfo{person}{Ming Zhong}, \bibinfo{person}{Jiaming Shen}, \bibinfo{person}{Yunyi Zhang}, \bibinfo{person}{Chao Zhang}, {and} \bibinfo{person}{Jiawei Han}.} \bibinfo{year}{2023}\natexlab{}.
\newblock \showarticletitle{Unsupervised Event Chain Mining from Multiple Documents}. In \bibinfo{booktitle}{\emph{{WWW}}}. \bibinfo{publisher}{{ACM}}, \bibinfo{pages}{1948--1959}.
\newblock


\bibitem[Jin et~al\mbox{.}(2021)]%
        {ForecastQA}
\bibfield{author}{\bibinfo{person}{Woojeong Jin}, \bibinfo{person}{Rahul Khanna}, \bibinfo{person}{Suji Kim}, \bibinfo{person}{Dong{-}Ho Lee}, \bibinfo{person}{Fred Morstatter}, \bibinfo{person}{Aram Galstyan}, {and} \bibinfo{person}{Xiang Ren}.} \bibinfo{year}{2021}\natexlab{}.
\newblock \showarticletitle{ForecastQA: {A} Question Answering Challenge for Event Forecasting with Temporal Text Data}. In \bibinfo{booktitle}{\emph{{ACL/IJCNLP} {(1)}}}. \bibinfo{publisher}{Association for Computational Linguistics}, \bibinfo{pages}{4636--4650}.
\newblock


\bibitem[Jin et~al\mbox{.}(2020)]%
        {RENET}
\bibfield{author}{\bibinfo{person}{Woojeong Jin}, \bibinfo{person}{Meng Qu}, \bibinfo{person}{Xisen Jin}, {and} \bibinfo{person}{Xiang Ren}.} \bibinfo{year}{2020}\natexlab{}.
\newblock \showarticletitle{Recurrent Event Network: Autoregressive Structure Inferenceover Temporal Knowledge Graphs}. In \bibinfo{booktitle}{\emph{{EMNLP} {(1)}}}. \bibinfo{publisher}{Association for Computational Linguistics}, \bibinfo{pages}{6669--6683}.
\newblock


\bibitem[Lee et~al\mbox{.}(2023)]%
        {GPT-NeoX-ICL}
\bibfield{author}{\bibinfo{person}{Dong{-}Ho Lee}, \bibinfo{person}{Kian Ahrabian}, \bibinfo{person}{Woojeong Jin}, \bibinfo{person}{Fred Morstatter}, {and} \bibinfo{person}{Jay Pujara}.} \bibinfo{year}{2023}\natexlab{}.
\newblock \showarticletitle{Temporal Knowledge Graph Forecasting Without Knowledge Using In-Context Learning}. In \bibinfo{booktitle}{\emph{{EMNLP}}}. \bibinfo{publisher}{Association for Computational Linguistics}, \bibinfo{pages}{544--557}.
\newblock


\bibitem[Lewis et~al\mbox{.}(2020)]%
        {RAG}
\bibfield{author}{\bibinfo{person}{Patrick S.~H. Lewis}, \bibinfo{person}{Ethan Perez}, \bibinfo{person}{Aleksandra Piktus}, \bibinfo{person}{Fabio Petroni}, \bibinfo{person}{Vladimir Karpukhin}, \bibinfo{person}{Naman Goyal}, \bibinfo{person}{Heinrich K{\"{u}}ttler}, \bibinfo{person}{Mike Lewis}, \bibinfo{person}{Wen{-}tau Yih}, \bibinfo{person}{Tim Rockt{\"{a}}schel}, \bibinfo{person}{Sebastian Riedel}, {and} \bibinfo{person}{Douwe Kiela}.} \bibinfo{year}{2020}\natexlab{}.
\newblock \showarticletitle{Retrieval-Augmented Generation for Knowledge-Intensive {NLP} Tasks}. In \bibinfo{booktitle}{\emph{NeurIPS}}.
\newblock


\bibitem[Li et~al\mbox{.}(2024b)]%
        {li2024multimodal}
\bibfield{author}{\bibinfo{person}{Chunyuan Li}, \bibinfo{person}{Zhe Gan}, \bibinfo{person}{Zhengyuan Yang}, \bibinfo{person}{Jianwei Yang}, \bibinfo{person}{Linjie Li}, \bibinfo{person}{Lijuan Wang}, \bibinfo{person}{Jianfeng Gao}, {et~al\mbox{.}}} \bibinfo{year}{2024}\natexlab{b}.
\newblock \showarticletitle{Multimodal foundation models: From specialists to general-purpose assistants}.
\newblock \bibinfo{journal}{\emph{Foundations and Trends{\textregistered} in Computer Graphics and Vision}} \bibinfo{volume}{16}, \bibinfo{number}{1-2} (\bibinfo{year}{2024}), \bibinfo{pages}{1--214}.
\newblock


\bibitem[Li et~al\mbox{.}(2023)]%
        {li2023your}
\bibfield{author}{\bibinfo{person}{Haoxuan Li}, \bibinfo{person}{Yi Bin}, \bibinfo{person}{Junrong Liao}, \bibinfo{person}{Yang Yang}, {and} \bibinfo{person}{Heng~Tao Shen}.} \bibinfo{year}{2023}\natexlab{}.
\newblock \showarticletitle{Your negative may not be true negative: Boosting image-text matching with false negative elimination}. In \bibinfo{booktitle}{\emph{Proceedings of the 31st ACM International Conference on Multimedia}}. \bibinfo{pages}{924--934}.
\newblock


\bibitem[Li et~al\mbox{.}(2024a)]%
        {li2024focusing}
\bibfield{author}{\bibinfo{person}{Jun Li}, \bibinfo{person}{Yi Bin}, \bibinfo{person}{Liang Peng}, \bibinfo{person}{Yang Yang}, \bibinfo{person}{Yangyang Li}, \bibinfo{person}{Hao Jin}, {and} \bibinfo{person}{Zi Huang}.} \bibinfo{year}{2024}\natexlab{a}.
\newblock \showarticletitle{Focusing on Relevant Responses for Multi-modal Rumor Detection}.
\newblock \bibinfo{journal}{\emph{IEEE Transactions on Knowledge and Data Engineering}} (\bibinfo{year}{2024}).
\newblock


\bibitem[Li et~al\mbox{.}(2022)]%
        {CLIP-Event}
\bibfield{author}{\bibinfo{person}{Manling Li}, \bibinfo{person}{Ruochen Xu}, \bibinfo{person}{Shuohang Wang}, \bibinfo{person}{Luowei Zhou}, \bibinfo{person}{Xudong Lin}, \bibinfo{person}{Chenguang Zhu}, \bibinfo{person}{Michael Zeng}, \bibinfo{person}{Heng Ji}, {and} \bibinfo{person}{Shih{-}Fu Chang}.} \bibinfo{year}{2022}\natexlab{}.
\newblock \showarticletitle{CLIP-Event: Connecting Text and Images with Event Structures}. In \bibinfo{booktitle}{\emph{{CVPR}}}. \bibinfo{publisher}{{IEEE}}, \bibinfo{pages}{16399--16408}.
\newblock


\bibitem[Li et~al\mbox{.}(2021)]%
        {REGCN}
\bibfield{author}{\bibinfo{person}{Zixuan Li}, \bibinfo{person}{Xiaolong Jin}, \bibinfo{person}{Wei Li}, \bibinfo{person}{Saiping Guan}, \bibinfo{person}{Jiafeng Guo}, \bibinfo{person}{Huawei Shen}, \bibinfo{person}{Yuanzhuo Wang}, {and} \bibinfo{person}{Xueqi Cheng}.} \bibinfo{year}{2021}\natexlab{}.
\newblock \showarticletitle{Temporal Knowledge Graph Reasoning Based on Evolutional Representation Learning}. In \bibinfo{booktitle}{\emph{{SIGIR}}}. \bibinfo{publisher}{{ACM}}, \bibinfo{pages}{408--417}.
\newblock


\bibitem[Liang et~al\mbox{.}(2024)]%
        {Foundation_TS}
\bibfield{author}{\bibinfo{person}{Yuxuan Liang}, \bibinfo{person}{Haomin Wen}, \bibinfo{person}{Yuqi Nie}, \bibinfo{person}{Yushan Jiang}, \bibinfo{person}{Ming Jin}, \bibinfo{person}{Dongjin Song}, \bibinfo{person}{Shirui Pan}, {and} \bibinfo{person}{Qingsong Wen}.} \bibinfo{year}{2024}\natexlab{}.
\newblock \showarticletitle{Foundation Models for Time Series Analysis: A Tutorial and Survey}.
\newblock \bibinfo{journal}{\emph{arXiv preprint arXiv:2403.14735}} (\bibinfo{year}{2024}).
\newblock


\bibitem[Liao et~al\mbox{.}(2023)]%
        {GENTKG}
\bibfield{author}{\bibinfo{person}{Ruotong Liao}, \bibinfo{person}{Xu Jia}, \bibinfo{person}{Yunpu Ma}, {and} \bibinfo{person}{Volker Tresp}.} \bibinfo{year}{2023}\natexlab{}.
\newblock \showarticletitle{GenTKG: Generative Forecasting on Temporal Knowledge Graph}.
\newblock \bibinfo{journal}{\emph{CoRR}}  \bibinfo{volume}{abs/2310.07793} (\bibinfo{year}{2023}).
\newblock


\bibitem[Liu et~al\mbox{.}(2024)]%
        {llava}
\bibfield{author}{\bibinfo{person}{Haotian Liu}, \bibinfo{person}{Chunyuan Li}, \bibinfo{person}{Qingyang Wu}, {and} \bibinfo{person}{Yong~Jae Lee}.} \bibinfo{year}{2024}\natexlab{}.
\newblock \showarticletitle{Visual instruction tuning}. In \bibinfo{booktitle}{\emph{NeurIPS}}.
\newblock


\bibitem[Liu(2022)]%
        {LlamaIndex}
\bibfield{author}{\bibinfo{person}{Jerry Liu}.} \bibinfo{year}{2022}\natexlab{}.
\newblock \bibinfo{booktitle}{\emph{{LlamaIndex}}}.
\newblock
\urldef\tempurl%
\url{https://doi.org/10.5281/zenodo.1234}
\showDOI{\tempurl}


\bibitem[Luo et~al\mbox{.}(2024)]%
        {Chain-of-History}
\bibfield{author}{\bibinfo{person}{Ruilin Luo}, \bibinfo{person}{Tianle Gu}, \bibinfo{person}{Haoling Li}, \bibinfo{person}{Junzhe Li}, \bibinfo{person}{Zicheng Lin}, \bibinfo{person}{Jiayi Li}, {and} \bibinfo{person}{Yujiu Yang}.} \bibinfo{year}{2024}\natexlab{}.
\newblock \showarticletitle{Chain of History: Learning and Forecasting with LLMs for Temporal Knowledge Graph Completion}.
\newblock \bibinfo{journal}{\emph{CoRR}}  \bibinfo{volume}{abs/2401.06072} (\bibinfo{year}{2024}).
\newblock


\bibitem[Lv et~al\mbox{.}(2020)]%
        {scriptLearning}
\bibfield{author}{\bibinfo{person}{Shangwen Lv}, \bibinfo{person}{Fuqing Zhu}, {and} \bibinfo{person}{Songlin Hu}.} \bibinfo{year}{2020}\natexlab{}.
\newblock \showarticletitle{Integrating external event knowledge for script learning}. In \bibinfo{booktitle}{\emph{Proceedings of the 28th International Conference on Computational Linguistics}}. \bibinfo{pages}{306--315}.
\newblock


\bibitem[Ma et~al\mbox{.}(2023a)]%
        {SeCoGD}
\bibfield{author}{\bibinfo{person}{Yunshan Ma}, \bibinfo{person}{Chenchen Ye}, \bibinfo{person}{Zijian Wu}, \bibinfo{person}{Xiang Wang}, \bibinfo{person}{Yixin Cao}, {and} \bibinfo{person}{Tat{-}Seng Chua}.} \bibinfo{year}{2023}\natexlab{a}.
\newblock \showarticletitle{Context-aware Event Forecasting via Graph Disentanglement}. In \bibinfo{booktitle}{\emph{{KDD}}}. \bibinfo{publisher}{{ACM}}, \bibinfo{pages}{1643--1652}.
\newblock


\bibitem[Ma et~al\mbox{.}(2023b)]%
        {LoGo}
\bibfield{author}{\bibinfo{person}{Yunshan Ma}, \bibinfo{person}{Chenchen Ye}, \bibinfo{person}{Zijian Wu}, \bibinfo{person}{Xiang Wang}, \bibinfo{person}{Yixin Cao}, \bibinfo{person}{Liang Pang}, {and} \bibinfo{person}{Tat{-}Seng Chua}.} \bibinfo{year}{2023}\natexlab{b}.
\newblock \showarticletitle{Structured, Complex and Time-complete Temporal Event Forecasting}.
\newblock \bibinfo{journal}{\emph{CoRR}}  \bibinfo{volume}{abs/2312.01052} (\bibinfo{year}{2023}).
\newblock


\bibitem[Morstatter(2021)]%
        {RCT_B}
\bibfield{author}{\bibinfo{person}{Fred Morstatter}.} \bibinfo{year}{2021}\natexlab{}.
\newblock \showarticletitle{{RCT-B}}.
\newblock  (\bibinfo{year}{2021}).
\newblock
\urldef\tempurl%
\url{https://doi.org/10.7910/DVN/ROTHFT}
\showDOI{\tempurl}


\bibitem[Ning et~al\mbox{.}(2020)]%
        {torque}
\bibfield{author}{\bibinfo{person}{Qiang Ning}, \bibinfo{person}{Hao Wu}, \bibinfo{person}{Rujun Han}, \bibinfo{person}{Nanyun Peng}, \bibinfo{person}{Matt Gardner}, {and} \bibinfo{person}{Dan Roth}.} \bibinfo{year}{2020}\natexlab{}.
\newblock \showarticletitle{{TORQUE}: A Reading Comprehension Dataset of Temporal Ordering Questions}. In \bibinfo{booktitle}{\emph{{EMNLP}}}. \bibinfo{pages}{1158--1172}.
\newblock
\urldef\tempurl%
\url{https://doi.org/10.18653/v1/2020.emnlp-main.88}
\showDOI{\tempurl}


\bibitem[Park et~al\mbox{.}(2022)]%
        {EvoKG}
\bibfield{author}{\bibinfo{person}{Namyong Park}, \bibinfo{person}{Fuchen Liu}, \bibinfo{person}{Purvanshi Mehta}, \bibinfo{person}{Dana Cristofor}, \bibinfo{person}{Christos Faloutsos}, {and} \bibinfo{person}{Yuxiao Dong}.} \bibinfo{year}{2022}\natexlab{}.
\newblock \showarticletitle{EvoKG: Jointly Modeling Event Time and Network Structure for Reasoning over Temporal Knowledge Graphs}. In \bibinfo{booktitle}{\emph{{WSDM}}}. \bibinfo{publisher}{{ACM}}, \bibinfo{pages}{794--803}.
\newblock


\bibitem[Robertson et~al\mbox{.}(2009)]%
        {BM25}
\bibfield{author}{\bibinfo{person}{Stephen Robertson}, \bibinfo{person}{Hugo Zaragoza}, {et~al\mbox{.}}} \bibinfo{year}{2009}\natexlab{}.
\newblock \showarticletitle{The probabilistic relevance framework: BM25 and beyond}.
\newblock \bibinfo{journal}{\emph{Foundations and Trends{\textregistered} in Information Retrieval}} \bibinfo{volume}{3}, \bibinfo{number}{4} (\bibinfo{year}{2009}), \bibinfo{pages}{333--389}.
\newblock


\bibitem[Schlichtkrull et~al\mbox{.}(2018)]%
        {RGCN}
\bibfield{author}{\bibinfo{person}{Michael~Sejr Schlichtkrull}, \bibinfo{person}{Thomas~N. Kipf}, \bibinfo{person}{Peter Bloem}, \bibinfo{person}{Rianne van~den Berg}, \bibinfo{person}{Ivan Titov}, {and} \bibinfo{person}{Max Welling}.} \bibinfo{year}{2018}\natexlab{}.
\newblock \showarticletitle{Modeling Relational Data with Graph Convolutional Networks}. In \bibinfo{booktitle}{\emph{{ESWC}}} \emph{(\bibinfo{series}{Lecture Notes in Computer Science}, Vol.~\bibinfo{volume}{10843})}. \bibinfo{publisher}{Springer}, \bibinfo{pages}{593--607}.
\newblock


\bibitem[Shang et~al\mbox{.}(2019)]%
        {ConvTransE}
\bibfield{author}{\bibinfo{person}{Chao Shang}, \bibinfo{person}{Yun Tang}, \bibinfo{person}{Jing Huang}, \bibinfo{person}{Jinbo Bi}, \bibinfo{person}{Xiaodong He}, {and} \bibinfo{person}{Bowen Zhou}.} \bibinfo{year}{2019}\natexlab{}.
\newblock \showarticletitle{End-to-End Structure-Aware Convolutional Networks for Knowledge Base Completion}. In \bibinfo{booktitle}{\emph{{AAAI}}}. \bibinfo{publisher}{{AAAI} Press}, \bibinfo{pages}{3060--3067}.
\newblock


\bibitem[Sun et~al\mbox{.}(2023)]%
        {ToG}
\bibfield{author}{\bibinfo{person}{Jiashuo Sun}, \bibinfo{person}{Chengjin Xu}, \bibinfo{person}{Lumingyuan Tang}, \bibinfo{person}{Saizhuo Wang}, \bibinfo{person}{Chen Lin}, \bibinfo{person}{Yeyun Gong}, \bibinfo{person}{Heung{-}Yeung Shum}, {and} \bibinfo{person}{Jian Guo}.} \bibinfo{year}{2023}\natexlab{}.
\newblock \showarticletitle{Think-on-Graph: Deep and Responsible Reasoning of Large Language Model with Knowledge Graph}.
\newblock \bibinfo{journal}{\emph{CoRR}}  \bibinfo{volume}{abs/2307.07697} (\bibinfo{year}{2023}).
\newblock


\bibitem[Sun et~al\mbox{.}(2019)]%
        {RotatE}
\bibfield{author}{\bibinfo{person}{Zhiqing Sun}, \bibinfo{person}{Zhi{-}Hong Deng}, \bibinfo{person}{Jian{-}Yun Nie}, {and} \bibinfo{person}{Jian Tang}.} \bibinfo{year}{2019}\natexlab{}.
\newblock \showarticletitle{RotatE: Knowledge Graph Embedding by Relational Rotation in Complex Space}. In \bibinfo{booktitle}{\emph{{ICLR} (Poster)}}. \bibinfo{publisher}{OpenReview.net}.
\newblock


\bibitem[Tan et~al\mbox{.}(2023)]%
        {tempbench}
\bibfield{author}{\bibinfo{person}{Qingyu Tan}, \bibinfo{person}{Hwee~Tou Ng}, {and} \bibinfo{person}{Lidong Bing}.} \bibinfo{year}{2023}\natexlab{}.
\newblock \showarticletitle{Towards Benchmarking and Improving the Temporal Reasoning Capability of Large Language Models}. In \bibinfo{booktitle}{\emph{{ACL}}}. \bibinfo{publisher}{Association for Computational Linguistics}, \bibinfo{pages}{14820--14835}.
\newblock
\urldef\tempurl%
\url{https://doi.org/10.18653/v1/2023.acl-long.828}
\showDOI{\tempurl}


\bibitem[Team et~al\mbox{.}(2023)]%
        {gemini}
\bibfield{author}{\bibinfo{person}{Gemini Team}, \bibinfo{person}{Rohan Anil}, \bibinfo{person}{Sebastian Borgeaud}, \bibinfo{person}{Yonghui Wu}, \bibinfo{person}{Jean-Baptiste Alayrac}, \bibinfo{person}{Jiahui Yu}, \bibinfo{person}{Radu Soricut}, \bibinfo{person}{Johan Schalkwyk}, \bibinfo{person}{Andrew~M Dai}, \bibinfo{person}{Anja Hauth}, {et~al\mbox{.}}} \bibinfo{year}{2023}\natexlab{}.
\newblock \showarticletitle{Gemini: a family of highly capable multimodal models}.
\newblock \bibinfo{journal}{\emph{arXiv preprint arXiv:2312.11805}} (\bibinfo{year}{2023}).
\newblock


\bibitem[Tong et~al\mbox{.}(2020)]%
        {DRMM}
\bibfield{author}{\bibinfo{person}{Meihan Tong}, \bibinfo{person}{Shuai Wang}, \bibinfo{person}{Yixin Cao}, \bibinfo{person}{Bin Xu}, \bibinfo{person}{Juanzi Li}, \bibinfo{person}{Lei Hou}, {and} \bibinfo{person}{Tat{-}Seng Chua}.} \bibinfo{year}{2020}\natexlab{}.
\newblock \showarticletitle{Image Enhanced Event Detection in News Articles}. In \bibinfo{booktitle}{\emph{{AAAI}}}. \bibinfo{publisher}{{AAAI} Press}, \bibinfo{pages}{9040--9047}.
\newblock


\bibitem[Touvron et~al\mbox{.}(2023)]%
        {LLaMA}
\bibfield{author}{\bibinfo{person}{Hugo Touvron}, \bibinfo{person}{Thibaut Lavril}, \bibinfo{person}{Gautier Izacard}, \bibinfo{person}{Xavier Martinet}, \bibinfo{person}{Marie{-}Anne Lachaux}, \bibinfo{person}{Timoth{\'{e}}e Lacroix}, \bibinfo{person}{Baptiste Rozi{\`{e}}re}, \bibinfo{person}{Naman Goyal}, \bibinfo{person}{Eric Hambro}, \bibinfo{person}{Faisal Azhar}, \bibinfo{person}{Aur{\'{e}}lien Rodriguez}, \bibinfo{person}{Armand Joulin}, \bibinfo{person}{Edouard Grave}, {and} \bibinfo{person}{Guillaume Lample}.} \bibinfo{year}{2023}\natexlab{}.
\newblock \showarticletitle{LLaMA: Open and Efficient Foundation Language Models}.
\newblock \bibinfo{journal}{\emph{CoRR}}  \bibinfo{volume}{abs/2302.13971} (\bibinfo{year}{2023}).
\newblock


\bibitem[Wang and Zhao(2023)]%
        {tram}
\bibfield{author}{\bibinfo{person}{Yuqing Wang} {and} \bibinfo{person}{Yun Zhao}.} \bibinfo{year}{2023}\natexlab{}.
\newblock \showarticletitle{TRAM: Benchmarking Temporal Reasoning for Large Language Models}.
\newblock  (\bibinfo{year}{2023}).
\newblock
\showeprint[arxiv]{2310.00835}


\bibitem[Xu et~al\mbox{.}(2023)]%
        {PPT}
\bibfield{author}{\bibinfo{person}{Wenjie Xu}, \bibinfo{person}{Ben Liu}, \bibinfo{person}{Miao Peng}, \bibinfo{person}{Xu Jia}, {and} \bibinfo{person}{Min Peng}.} \bibinfo{year}{2023}\natexlab{}.
\newblock \showarticletitle{Pre-trained Language Model with Prompts for Temporal Knowledge Graph Completion}. In \bibinfo{booktitle}{\emph{{ACL} (Findings)}}. \bibinfo{publisher}{Association for Computational Linguistics}, \bibinfo{pages}{7790--7803}.
\newblock


\bibitem[Yang et~al\mbox{.}(2015)]%
        {DistMult}
\bibfield{author}{\bibinfo{person}{Bishan Yang}, \bibinfo{person}{Wen{-}tau Yih}, \bibinfo{person}{Xiaodong He}, \bibinfo{person}{Jianfeng Gao}, {and} \bibinfo{person}{Li Deng}.} \bibinfo{year}{2015}\natexlab{}.
\newblock \showarticletitle{Embedding Entities and Relations for Learning and Inference in Knowledge Bases}. In \bibinfo{booktitle}{\emph{{ICLR} (Poster)}}.
\newblock


\bibitem[Ye et~al\mbox{.}(2024)]%
        {ye2024mirai}
\bibfield{author}{\bibinfo{person}{Chenchen Ye}, \bibinfo{person}{Ziniu Hu}, \bibinfo{person}{Yihe Deng}, \bibinfo{person}{Zijie Huang}, \bibinfo{person}{Mingyu~Derek Ma}, \bibinfo{person}{Yanqiao Zhu}, {and} \bibinfo{person}{Wei Wang}.} \bibinfo{year}{2024}\natexlab{}.
\newblock \showarticletitle{MIRAI: Evaluating LLM Agents for Event Forecasting}.
\newblock \bibinfo{journal}{\emph{arXiv preprint arXiv:2407.01231}} (\bibinfo{year}{2024}).
\newblock


\bibitem[Zhang and Choi(2021)]%
        {situatedqa}
\bibfield{author}{\bibinfo{person}{Michael Zhang} {and} \bibinfo{person}{Eunsol Choi}.} \bibinfo{year}{2021}\natexlab{}.
\newblock \showarticletitle{SituatedQA: Incorporating Extra-Linguistic Contexts into QA}. In \bibinfo{booktitle}{\emph{{EMNLP}}}.
\newblock


\bibitem[Zhang et~al\mbox{.}(2022)]%
        {opt}
\bibfield{author}{\bibinfo{person}{Susan Zhang}, \bibinfo{person}{Stephen Roller}, \bibinfo{person}{Naman Goyal}, \bibinfo{person}{Mikel Artetxe}, \bibinfo{person}{Moya Chen}, \bibinfo{person}{Shuohui Chen}, \bibinfo{person}{Christopher Dewan}, \bibinfo{person}{Mona Diab}, \bibinfo{person}{Xian Li}, \bibinfo{person}{Xi~Victoria Lin}, {et~al\mbox{.}}} \bibinfo{year}{2022}\natexlab{}.
\newblock \showarticletitle{Opt: Open pre-trained transformer language models}.
\newblock \bibinfo{journal}{\emph{arXiv preprint arXiv:2205.01068}} (\bibinfo{year}{2022}).
\newblock


\bibitem[Zhang et~al\mbox{.}(2024)]%
        {zhang2024analyzing}
\bibfield{author}{\bibinfo{person}{Zhihan Zhang}, \bibinfo{person}{Yixin Cao}, \bibinfo{person}{Chenchen Ye}, \bibinfo{person}{Yunshan Ma}, \bibinfo{person}{Lizi Liao}, {and} \bibinfo{person}{Tat-Seng Chua}.} \bibinfo{year}{2024}\natexlab{}.
\newblock \showarticletitle{Analyzing Temporal Complex Events with Large Language Models? A Benchmark towards Temporal, Long Context Understanding}.
\newblock \bibinfo{journal}{\emph{arXiv preprint arXiv:2406.02472}} (\bibinfo{year}{2024}).
\newblock


\bibitem[Zhao(2021)]%
        {zhao2021event}
\bibfield{author}{\bibinfo{person}{Liang Zhao}.} \bibinfo{year}{2021}\natexlab{}.
\newblock \showarticletitle{Event prediction in the big data era: A systematic survey}.
\newblock \bibinfo{journal}{\emph{ACM Computing Surveys (CSUR)}} \bibinfo{volume}{54}, \bibinfo{number}{5} (\bibinfo{year}{2021}), \bibinfo{pages}{1--37}.
\newblock


\bibitem[Zhou et~al\mbox{.}(2019)]%
        {going}
\bibfield{author}{\bibinfo{person}{Ben Zhou}, \bibinfo{person}{Daniel Khashabi}, \bibinfo{person}{Qiang Ning}, {and} \bibinfo{person}{Dan Roth}.} \bibinfo{year}{2019}\natexlab{}.
\newblock \showarticletitle{{``}Going on a vacation{''} takes longer than {``}Going for a walk{''}: A Study of Temporal Commonsense Understanding}. In \bibinfo{booktitle}{\emph{{EMNLP}}}. \bibinfo{pages}{3363--3369}.
\newblock
\urldef\tempurl%
\url{https://doi.org/10.18653/v1/D19-1332}
\showDOI{\tempurl}


\end{thebibliography}
